\documentclass[a4paper,fleqn,usenatbib]{mnras}
\usepackage{newtxtext,newtxmath}

\usepackage[T1]{fontenc}
\usepackage{ae,aecompl}

\usepackage{amsmath,graphicx,amssymb}
\usepackage{natbib}
\usepackage{pst-eps}
\bibpunct{(}{)}{,}{a}{}{,}

\newcommand{\hzav}[1]{\left[#1\right]}

\newlength\staretab

\newcommand{\tuv}{\ensuremath{T_\text{UV}}}
\newcommand\de{\text{d}}

\newcommand\kms{\ensuremath{\text{km}\,\text{s}^{-1}}}

\newcommand\hvezda{hvezda}
\newcommand\ergs{\ensuremath{\text{erg}\,\text{s}^{-1}\,\text{cm}^{-2}}}
\newcommand\ergsa{\ensuremath{\text{erg}\,\text{s}^{-1}\,\text{cm}^{-2}\,
\text{\AA}^{-1}}}

\title[The ultraviolet study of B{[e]} stars]{The ultraviolet study of B[e] stars: evidence for pulsations,
LBV-type variations, and processes in envelope}

\author[I.~Krti\v{c}kov\'a and J.~Krti\v{c}ka]{I.~Krti\v{c}kov\'a\thanks{E-mail: 15124@mail.muni.cz}
and J.~Krti\v{c}ka\\
Department of Theoretical Physics and Astrophysics,
Masaryk University, Kotl\'a\v rsk\' a 2, CZ-611\,37 Brno, Czech
Republic}

\date{Accepted XXX. Received YYY; in original form ZZZ}

\pubyear{2015}

\begin{document}
\label{firstpage}
\pagerange{\pageref{firstpage}--\pageref{lastpage}}
\maketitle

\begin{abstract}
Stars with B[e] phenomenon comprise a very diverse group of objects in a
different evolutionary status. These objects show common spectral
characteristics, including presence of Balmer lines in emission, forbidden
lines, and strong infrared excess due to the dust. The observations of emission
lines indicate the illumination by ultraviolet ionizing source, which is a key
part to understand the elusive nature of these objects. We study the
ultraviolet variability of many B[e] stars to specify the geometry of the
circumstellar environment and its variability. We analyse massive
hot B[e] stars from our Galaxy and from Magellanic Clouds.
We study the ultraviolet broad-band variability derived from the flux-calibrated
data. We determine variations of individual
lines and its correlation with the total flux variability. We detected 
variability of the spectral energy distribution and of the line profiles.
The variability has several sources of origin, including the light
absorption by the disk, pulsations, LBV-type variations, and eclipses in the
case of binaries. The stellar radiation of most of B[e] stars is heavily
obscured by circumstellar material. This suggests that the circumstellar
material is not present only in the disk but also above its plane. The flux and
line variability is consistent with a two-component model of circumstellar
environment composed of the dense disk and ionized envelope. The observations of
B[e] supergiants show that many of these stars have nearly the same luminosity
of about $1.9\times10^{5}\,L_\odot$ and similar effective temperature.
\end{abstract}


\section{Introduction}

B[e] stars form a very diverse group of stars that share common spectroscopic
properties, however their astrophysical nature is very different. Among the
typical spectroscopic properties of B[e] stars belong strong Balmer lines in
emission, presence of low excitation permited emission lines as well as
forbidden emission lines, and infrared excess \citep{alles}. Despite their
spectral similarities, the astrophysical nature of stars showing B[e] phenomenon
is diverse. The group of B[e] stars contains stars in different evolutionary
stages (pre-main sequence B[e] stars vs.~B[e] supergiants), stars with different
initial mass (massive B[e] supergiants vs.~compact planetary nebulae), and
binary status (including symbiotic binaries). Besides these classes of stars
showing B[e] phenomenon summarized by \citet{labe} also presumably unevolved
B[e] stars may appear \citep{mirotrida}.

The variability observed in the optical photometry (e.g.,
\citealt{sitovygraf,zimajda,pohlasterky}) and spectroscopy
\citep[e.g.,][]{borecferdapul,polstar,kucerka} indicates a possibility of
ultraviolet (UV) variability. Although the UV flux variability was indeed found
in some B[e] stars \citep{odpadnici,shorec,sitovygraf}, there is no detailed
systematic study of UV variability of B[e] stars according to our knowledge.

The study of UV variability of B[e] stars is especially relevant because hot
stars emit most of their radiation in the UV domain. Consequently, the UV
variability may shed some light on the elusive nature of many of these objects.
The circumstellar environment of B[e] stars is combined from an ionized hot
envelope and cool dusty material \citep{shorsan}. While the presence of ionized
envelope can be naturally explained by radiatively driven stellar wind, which is
common in luminous stars of spectral type B \citep{vikola,zelib,mark}, the
origin of a cool dust containing envelope is unclear. The cool envelope is
likely shaped in the form of a disk \citep{dalsiziki,schula} whose inner parts
may be relatively hot and shield the remaining cooler parts
\citep[e.g.,][]{kraul,zshg}.

The geometry of the circumstellar environment may therefore to some extent
resemble classical Be stars \citep[see][for a review of Be phenomenon]{ricam}.
Classical Be stars may contain a viscous disk that is fed as a result of angular
momentum loss from near critically rotating star
\citep{los91,sapporo,kom,petrk}. The origin of the disk is still a matter of
debate in Be stars as well as in B[e] stars. The near critical rotation in these
stars may possibly be induced either by evolutionary spin up \citep{granada} or
by a binary interaction \citep{botu}. The later hypothesis may be more likely in
B[e] supergiants because the evolutionary models show strong decrease of the
rotational velocity in the supergiant phase \citep{sitrot}.

The merger origin of B[e] stars was proposed by, e.g., \citet{pods}. The flow
during such event is complex and may involve outflows as well as inflows
\citep[e.g.,][]{pycha}. The merger scenario may be supported by the presence of
extended envelopes that appear around some B[e] stars (e.g., HD~34664,
\citealt{chuchen}, HD~38489, \citealt{kasar}). Some B[e] stars are spectroscopic
binaries \citep[e.g.,][]{djurak,brezina}. 

In the following paper, we analyse the archival IUE data available for B[e]
stars to better understand the geometry of the circumstellar environment around
these stars and to describe its variability in detail. Because the group of
B[e] stars is very diverse, we focus on the stars that may share some common
characteristics and exclude pre-main sequence stars, planetary nebulae, and
symbiotic binaries from our analysis. We study mostly unevolved and evolved
hot massive stars that display the B[e] phenomenon.

\section{Methods}
\label{metody}

We downloaded the far-UV 1150--1900~\AA\ (SWP camera) and near-UV
2000--3300~\AA\ (LWP and LWR cameras) low-dispersion large-aperture
fluxes of B[e] stars from
the INES database using the SPLAT-VO package \citep{splat,pitr}. We selected only
B[e] stars with multiple IUE observations that are suitable for the analysis of
variability. The list of all used IUE observations is given in
Appendix.

\begin{figure}
\centering
\resizebox{\hsize}{!}{\includegraphics{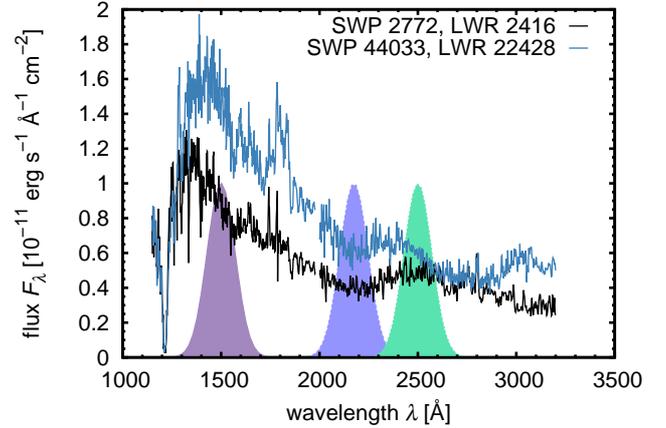}}
\caption{Example of the analysis of the broad-band variations of HD~45677. We
plot the flux distribution for two different epochs (black and blue solid
curves). Overplotted are individual adopted bandpass filters (not to scale). The
change of broad-band fluxes is about 60\,\% in fluxes $F_{1500}$ and $F_{2175}$
and about 20\,\% in $F_{2500}$ (see Table~\ref{hd45677tab}).}
\label{metodaj}
\end{figure}

Motivated by the success in detection and modelling of the flux
variability of chemically peculiar stars from IUE spectra
\citep[e.g.,][]{mycuvir,myteta}, we selected the same approach to study the
variability of B[e] stars. We constructed the broad-band fluxes
\begin{equation}
F_c=\int_0^{\infty}\mathit\Phi_c(\lambda) F(\lambda)\,\de\lambda
\end{equation}
from the UV fluxes $F(\lambda)$ observed by IUE. Here
$\mathit\Phi_c(\lambda)=(\sqrt\pi\sigma)^{-1}
\exp\hzav{-(\lambda-c)^2/\sigma^2}$
is a Gauss function centered on the wavelength $c$. The central wavelengths of
individual Gauss filters were selected to describe the individual regions of UV
spectra. We focused on the regions without strong emission lines.
We selected $c=1500$\,\AA\ to describe the flux in far-UV regions,
$c=2175$\,\AA\ which selects the flux in the region of carbon opacity bump, and
$c=2500$\,\AA\ for near-UV region 
(see Fig.~\ref{metodaj}). The filters centered on different wavelength
regions allow us to study the
variations of temperature. Moreover, the regions of $c=1500$\,\AA\ and
$c=2500$\,\AA\ are affected by the interstellar reddening by nearly the same
amount \citep{fima}.
We selected $\sigma=100$\,\AA\ to cover a broader region of the stellar flux
distribution. However, our tests showed that the selection of
central wavelengths and the dispersion does not significantly affect final
results.
The variations
are plotted as a function of modified Julian date (MJD),
$\text{MJD}=\text{JD}-2\,400\,000$. The derived values together with their
mean uncertainties (in the same units) are given also in
Tables~\ref{sleps18}--\ref{slephd169515}. We used our results for the comparison
of individual fluxes and for the study of general relationships between them.

\begin{figure}
\centering
\resizebox{\hsize}{!}{\includegraphics{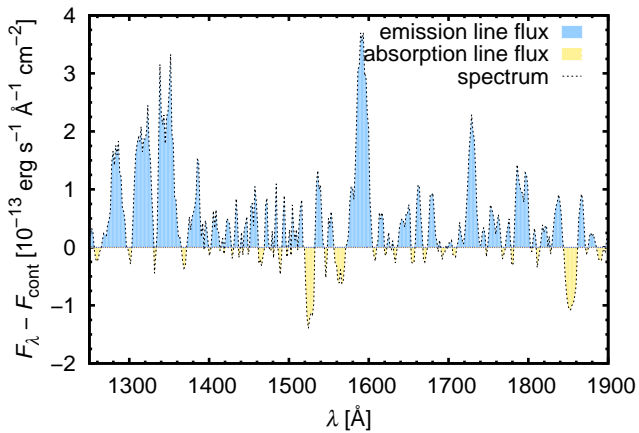}}
\caption{Example of the analysis of the line spectrum of HD~87643 (SWP~46913).
The dashed line denotes the observed spectrum from which the pseudocontinuum was
substracted. The total emission line flux is the total flux above the zero (blue
shaded area) and the total line absorption is the total flux below zero
(yellow shaded area).}
\label{metodai}
\end{figure}

The emission lines in the spectra of B[e] stars also clearly show variability
\citep[e.g.,][]{sanda,kucerka}.
However, their study is more problematic, because there is no clear
continuum in UV region,
and the true continuum may lie significantly above the anticipated level.
To overcome this problem, we constructed a pseudocontinuum in each spectrum and
substracted the pseudocontinuum flux from observed flux to derive the total flux
in individual lines (see Fig.~\ref{metodai}).
The pseudocontinuum was created by hand using the 
SPLAT-VO\footnote{http://star-www.dur.ac.uk/\~{}pdraper/splat/splat-vo/} package taking into
account the flux distribution of stars without B[e] phenomenon.
We used the spline function to connect the points corresponding to the
pseudocontinuum. In other words, a
similar procedure is typically applied as when normalizing a spectrum.
To describe the
variations of the emission and absorption lines as a whole, we integrated the 
flux above and below the pseudocontinuum in the region 1250--1900\,\AA\ to
derive the total flux in the emission lines $F_\text{em}$ and line absorption
$F_\text{abs}$ as shown in Fig.~\ref{metodai}. The fluxes in individual strong
lines and the total line fluxes are plotted in graphs and also listed in
Tables~\ref{sleps18}--\ref{slephd169515}. The uncertainties are given in
the same units as individual fluxes.

During the construction of the pseudocontinuum,
we focused our attention to a precise and
homogeneous processing of all spectra. The repeated processing of selected
spectra showed that we are able to reliably reproduce the derived results.
Consequently, the
line fluxes $F_\text{em}$ and $F_\text{abs}$
reflect the changes in the spectral properties of
individual stars and enable the mutual comparison between studied stars.
Moreover, the comparison with literature data shows that our emission line
fluxes agree typically within $10-20\,\%$ with that given in \citet{shorsan} for
HD~38489 (see~Table~\ref{hd38489tab}) and in \citet{shorenemapravdu} for
LHA~115-S~18 (Table~\ref{sleps18}). 

We also tested the dependence of the flux on the position angle of the slit to
reveal large nebula around the star or nearby stellar object. We have not
detected any such dependence for all stars except one. We found the dependence
of the flux on the position angle for LHA\,115-S\,18, however, we interpret this
dependence as coincidental.

Our additional aim was the determination of the temperature of the stellar
envelope. We fitted the observed spectra by theoretical spectral energy
distribution. We selected the spectra for which nearly simultaneous observations
are available in far-UV and near-UV regions. The spectra were fitted by ATLAS9
fluxes\footnote{http://www.oact.inaf.it/castelli/} \citep{kurat,casat}
calculated assuming LTE, which cover the studied temperature interval. We
selected solar chemical composition for the
Galactic stars and $\text{[M/H]}=-0.5$ for stars from the Magellanic
Clouds.
The spectra were attenuated by
the extinction curve $k(\lambda-V)$ of \citet{fima},
\begin{equation}
\label{fima}
F(\lambda)=F_0(\lambda)\,10^{-\alpha\, k(\lambda-V)},
\end{equation}
where $F_0(\lambda)$ is unattenuated flux and $\alpha$ is a free parameter of
the fit. The other fit parameters are the model effective temperature and
surface gravity. For normal stars without circumstellar envelope, the model
effective temperature corresponding to the best fit of the UV flux distribution
is equal to the effective temperature of the star. For the stars with opaque
circumstellar environment, the stellar effective temperature may be different,
and consequently we denote the temperature corresponding to the best fit of UV
spectra as \tuv. This temperature corresponds to the temperature of the
circumstellar envelope. Because the fluxes are rather insensitive to the surface
gravity, we do not provide its final value.
B[e] stars typically show infrared excess due to the dust. Consequently,
part of the dust absorption accounted by Eq.~\eqref{fima} may originate in
circumstellar environment of given star. However, because infrared dust emission
traces large volumes of the circumstellar medium, while the dust absorption is
given just by the dust particles intersecting the line of sight, we expect that
the interstellar contribution may dominate in Eq.~\eqref{fima}.
Anyway, both contributions are accounted for in Eq.~\eqref{fima} because
our approach does not assume any particular spatial distribution of the dust.
We fitted
just the normalized fluxes. 

\section{UV variability of individual stars}

We searched the IUE archive for the observations of B[e] stars
with multiple IUE spectra. Basic
properties of studied stars are given in Appendix~\ref{slepak} and are
summarized in Table~\ref{tabpar} including the temperature $T_\text{UV}$ derived
from the spectral energy distribution fit. We provide range of temperatures for
the stars that show variability of $T_\text{UV}$. We have not found any clear
variations of the dust absorption parameter $\alpha$ Eq.~\eqref{fima} in any
star. 

\begin{table}
\caption{Basic propertied and temperatures $T_\text{UV}$ (see
Sect.~\ref{metody}) of studied stars. Here ``sg''
denotes supergiant, ``b'' binary, ``eb'' eclipsing binary and ``H''
Herbig Ae/Be star.}
\label{tabpar}
\begin{center}
\begin{tabular}{l@{\hspace{2.5mm}}c@{\hspace{2.5mm}}c@{\hspace{2.5mm}}c}
\hline
Star & Location & Type & $T_\text{UV}$ [K]\\
\hline
LHA 115-S 18 (AzV 154) & SMC & & \\
LHA 115-S 65 (Sk 193)& SMC & sg & $9100\pm100$\\
LHA 120-S 12 (Sk -67 23)& LMC & sg & $18\,800\pm300$ \\
HD~34664 (LHA 120-S 22)& LMC & sg & $9\,500$ -- $10\,500$\\
HD 37974 (LHA 120-S 127)& LMC & sg & $12\,800\pm400$\\
HD 38489 (LHA\,120-S\,134)& LMC & sg & $16\,900\pm1000$\\
HD 45677 (FS CMa) & Gal. & & $14\,100$ -- $16\,300$\\
HD 50138 (V743 Mon) & Gal. & & $11\,700\pm200$\\
HD 87643 (V640 Car) & Gal. & b & $10\,900$ -- $11\,700$\\
HD 94878 (GG Car) & Gal. & eb & $27\,800\pm700$\\
HD 100546 (KR Mus) & Gal. & H & $12\,000\pm200$\\
HD 169515 (RY Sct)& Gal. & eb & $35\,000\pm1000$\\
\hline
\end{tabular}
\end{center}
\end{table}

\subsection{LHA 115-S 18}

\renewcommand\hvezda{LHA 115-S 18}

The star \hvezda\ shows a prominent UV flux variability
(Fig.~\ref{lha115s18}).
This variability is caused mostly by emission lines, whereas the continuum
varies only weakly \citep{shorenemapravdu}. In accordance with
\citet{shorenemapravdu}, we detected line variability in \ion{N}{v} 1245\,\AA,
\ion{O}{i} 1302\,\AA, \ion{C}{iv} 1550\,\AA, and \ion{He}{ii} 1640\,\AA\ lines
(Table~\ref{sleps18}). The total emission line flux and the total
line absorption flux vary (Table~\ref{sleps18}). The typical timescale of
the variability (about four years) is longer than the most prominent cycle of
optical variations, which is about 440~days \citep{uradnik}. We have not fitted
the spectrum of this star, because the continuum is relatively weak.


\subsection{LHA 115-S 65}

\renewcommand\hvezda{LHA 115-S 65}

All IUE observations of \hvezda\ were obtained in two different epochs.
Consequently, we calculated just mean values of individual fluxes in both epochs
(Table~\ref{iues65mag}). While the fluxes at
1500\,\AA\ and 2500\,\AA\ do not change within the errors, there may be a
decrease of the flux at 2175\,\AA. This may indicate increased dust
absorption by intervening circumstellar material.


\subsection{LHA 120-S 12}

\renewcommand\hvezda{LHA 120-S 12}

\begin{figure}
\centering
\resizebox{0.9\hsize}{!}{\includegraphics{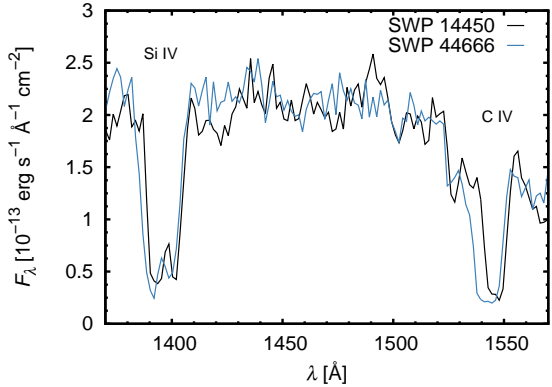}}
\caption{Variations of \hvezda\ wind line profiles.}
\label{lha120s12civ}
\end{figure}

The IUE observations of \hvezda\ are scarce. However, the wind line profiles
show variability. The most prominent is the increase of the wind terminal
velocity apparent in the \ion{C}{iv} 1548\,\AA\ line profile from
$1400\,\pm200\,\kms$ to $2300\,\pm200\,\kms$ (see Fig.~\ref{lha120s12civ})
during roughly 11\,years. Similar changes are present also in \ion{Si}{iv}
1393\,\AA\ line, albeit with a lesser extent. While the total line absorption
and the total emission line flux do not change substantially, the fluxes in
\ion{Si}{iv} doublet and in \ion{C}{iv} line  increase
(Table~\ref{sleps12}). If the fitted temperature $\tuv=18\,800\pm300\,\text{K}$
corresponds to the effective temperature of the star, then the change of the
wind terminal velocity can be explained as a result of bistability
\citep{bista}. This effect appears close to
$T_\text{eff}\approx20\,000\,\text{K}$ and leads to the modification of the wind
terminal velocity \citep{lsl,zelib} and possibly also of the mass-loss rate as a
result of sensitivity of wind ionization to the effective temperature of the
star \citep{vikolabis,petr}.

\subsection{HD 34664}

\renewcommand\hvezda{HD 34664}

\begin{figure}
\centering
\resizebox{0.9\hsize}{!}{\includegraphics{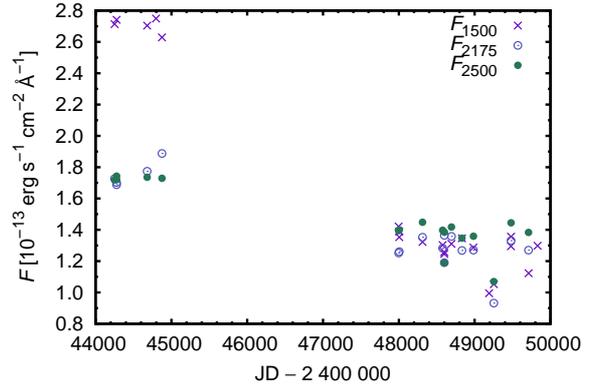}}
\caption{Variation of \hvezda\ broad-band fluxes with time.}
\label{hd34664prom}
\end{figure}

The flux variations of \hvezda\ in Fig.~\ref{hd34664prom} show a strong decrease
of the flux between MJD $45\,000$ and $48\,000$ detected by \citet{shorec}. The
decrease of the flux is stronger in far-UV band centered at 1500\,\AA\ than in
near-UV bands centered at 2175\,\AA\ and 2500\,\AA. This shows that the most
likely cause of the light variability is the change of the effective temperature
of the star. Such changes, which are typical for LBV S Doradus variables, were
detected in \hvezda\ by \citet{pohlasterky} and \citet{sterky} from optical
data. In accordance with this interpretation, \citet{pohlasterky} detected
optical reddening of this star over roughly the same period covered by IUE data.

\begin{figure}
\centering
\resizebox{0.9\hsize}{!}{\includegraphics{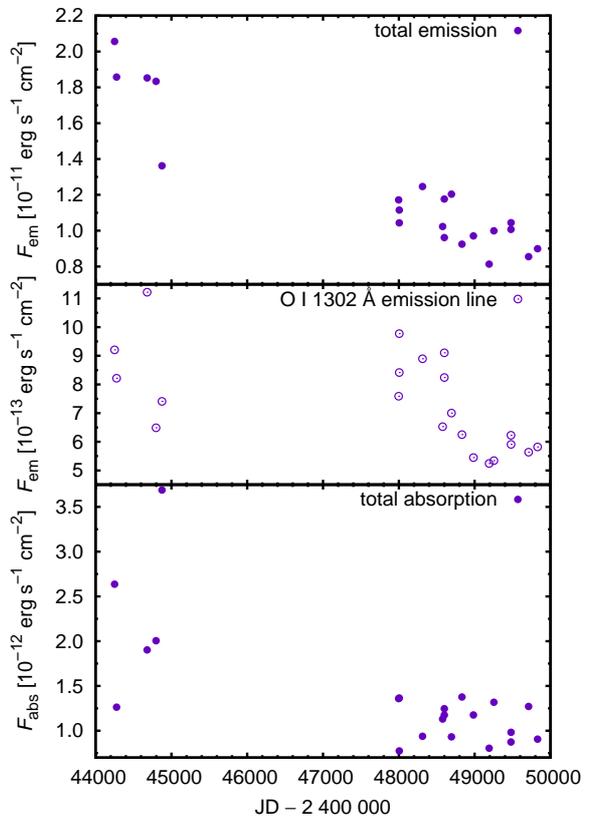}}
\caption{Variation of \hvezda\ total emission line flux ({\em upper panel}),
\ion{O}{i} 1302\,\AA\ emission line flux ({\em middle panel}), and the total line
absorption ({\em lower panel}) with time.}
\label{hd34664emabs}
\end{figure}

The variations of the total absorption and the total emission line fluxes in
Fig.~\ref{hd34664emabs} show linear decrease in both cases followed by the
period of constant flux in the case of the line absorption.
These variations of the total line fluxes correspond to the decrease of the
effective temperature of the star and are modified by the circumstellar
envelope. The most intriguing is the decrease of the emission line flux even
during the period of the constant flux after MJD 48\,000
(Fig.~\ref{hd34664emabs}, middle panel).
This may indicate that the regions of origin of emission and absorption lines
are different.

The temperature fitting yields a slow decrease from $\tuv=10\,500\pm200$~K
to $\tuv=9\,500\pm100$~K. This decrease corresponds to broad-band flux and
emission line flux variations. 

\subsection{HD 37974}

\renewcommand\hvezda{HD 37974}

The flux variations of \hvezda\ in Fig.~\ref{hd37974prom}
can be interpreted as the flux decrease during MJD 43\,000 -- 46\,000 and slight
brightening in the period MJD 48\,000 -- 50\,000. Additional variations may be
connected with complex long-term (hundreds of days) and short-time
(tens of days) optical light variability \citep{genster}.

\begin{figure}
\centering
\resizebox{0.9\hsize}{!}{\includegraphics{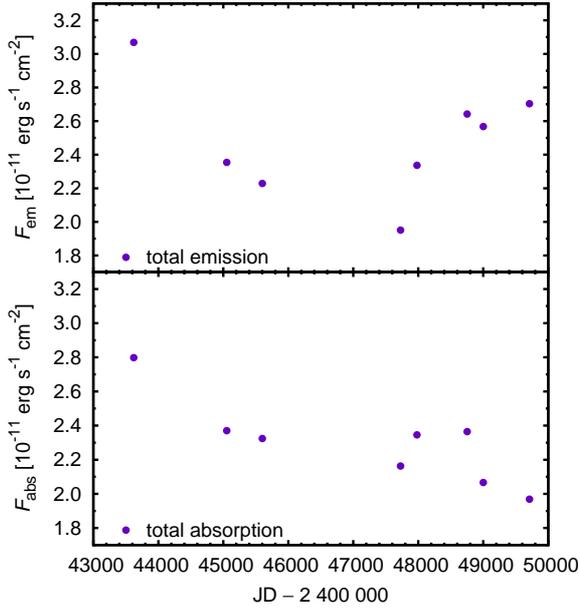}}
\caption{Variation of \hvezda\ total emission line flux ({\em upper panel})
and the total line absorption ({\em lower panel}) with time.}
\label{hd37974emabs}
\end{figure}

The variations of \ion{O}{i} 1302\,\AA\ emission line in
Fig.~\ref{hd37974prom} as well as
the total emission line flux in Fig.~\ref{hd37974emabs} show similar trends as
$F_{1500}$ fluxes, that is decrease during MJD 43\,000 -- 46\,000 and increase
of the later phases. The total line absorption decreased
during the whole observational period (Fig.~\ref{hd37974emabs}).


\subsection{HD 38489}

\renewcommand\hvezda{HD 38489}

\begin{figure}
\centering
\resizebox{0.9\hsize}{!}{\includegraphics{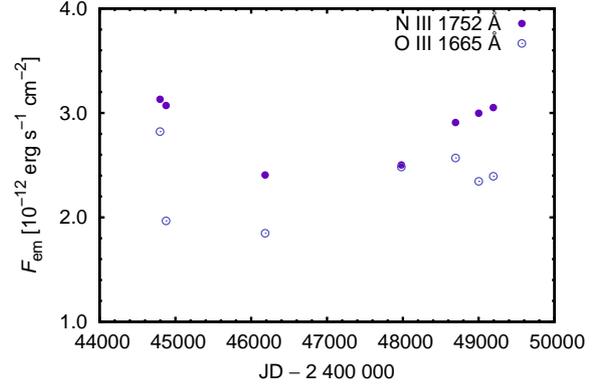}}
\caption{Variation of \hvezda\ fluxes in emission lines \ion{O}{iii}
1665\,\AA\ and \ion{N}{iii} 1752\,\AA\ lines with time.}
\label{hd38489em}
\end{figure}

The IUE spectra of \hvezda\ secured between MJD 44\,800 -- 49\,200 show nearly
constant flux\footnote{The observations at about MJD=49\,250 show flux dip, but
we regard this just as an instrumental feature, because a similar dip is present
in observations of HD~34664 (see Fig.~\ref{hd34664prom}).} \citep{shorsan} with
dispersion of about 2\,\%. Also the optical observations do not show any strong
variability \citep{ziki}. Despite this, the star shows emission line
variability. \ion{He}{ii} 1640\,\AA\ \citep{shorsan} emission line flux
increased during the period of IUE observation (Table~\ref{hd38489tab}).
On the
other hand, \ion{O}{iii} line at 1665\,\AA\ and \ion{N}{iii} line at 1752\,\AA\
show decrease followed by increase of the flux (see Fig.~\ref{hd38489em}). The
similarity of \ion{O}{iii} and \ion{N}{iii} emission line flux variations
supports the idea that these lines originate in the same region, which is
different from the region where \ion{He}{ii} line originates \citep{shorsan}.
Neither the total emission nor line absorption show a strong variability as a
result of differing trends in the variability of individual lines. 


\subsection{HD 45677}

\renewcommand\hvezda{HD 45677}


\begin{figure}
\centering
\resizebox{0.9\hsize}{!}{\includegraphics{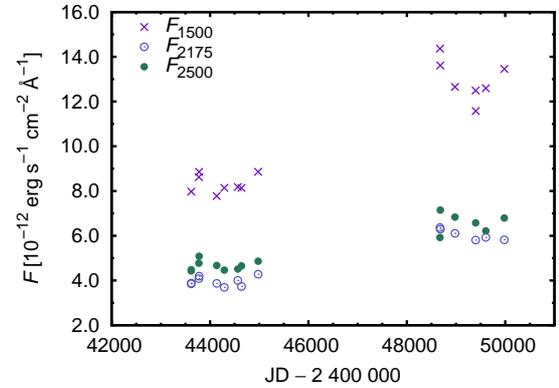}}
\caption{Variation of \hvezda\ broad-band fluxes with time.}
\label{hd45677prom}
\end{figure}

\begin{figure}
\centering
\resizebox{0.9\hsize}{!}{\includegraphics{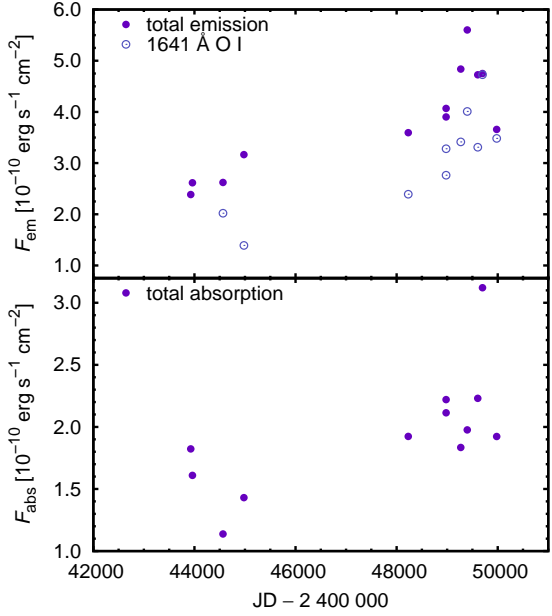}}
\caption{Variation of \hvezda\ fluxes in emission ({\em upper panel}) and
absorption ({\em lower panel}) lines with time. The emission flux in 1641\,\AA\
\ion{O}{i} line was artificially multiplied by ten to get the common scale of the
graph.} 
\label{hd45677emabs}
\end{figure}

The UV flux variations of \hvezda\ in Fig.~\ref{hd45677prom}, which show
increase of the flux from MJD of about 44\,000 to 48\,000, are correlated with
visual light variations \citep{sitovygraf}. Slightly decreasing trend between
MJD 48\,000 -- 50\,000 may be connected with regress from the visual light
maximum that appeared at about MJD 48\,700 \citep{sitovygraf}. The maximum is
visible also in the Hipparcos data \citep{ESA}.

The total emission flux, 1641\,\AA\ \ion{O}{i} emission line flux (see
Fig.~\ref{hd45677emabs}), and the 1745\,\AA\ \ion{N}{i}
\citep{sitnaodpad,hnedybus} line flux show a similar behaviour as the total
absorption and broad-band fluxes. This indicates that the corresponding
processes are connected.


The temperature $\tuv$
decreased from about $16\,300\pm500\,$K in MJD 43\,000 -- 45\,000 to
$14\,100\pm300\,$K in MJD 48\,000 -- 49\,000. The fluxes do not change
significantly below about 1300\,\AA, but significantly increase above 1300\,\AA\
in later epochs. Moreover, the flux maximum shifts from 1300\,\AA\ to about
1400\,\AA, causing the mentioned decrease of $\tuv$.

\subsection{HD 50138}

\renewcommand\hvezda{HD 50138}

\begin{figure}
\centering
\resizebox{0.9\hsize}{!}{\includegraphics{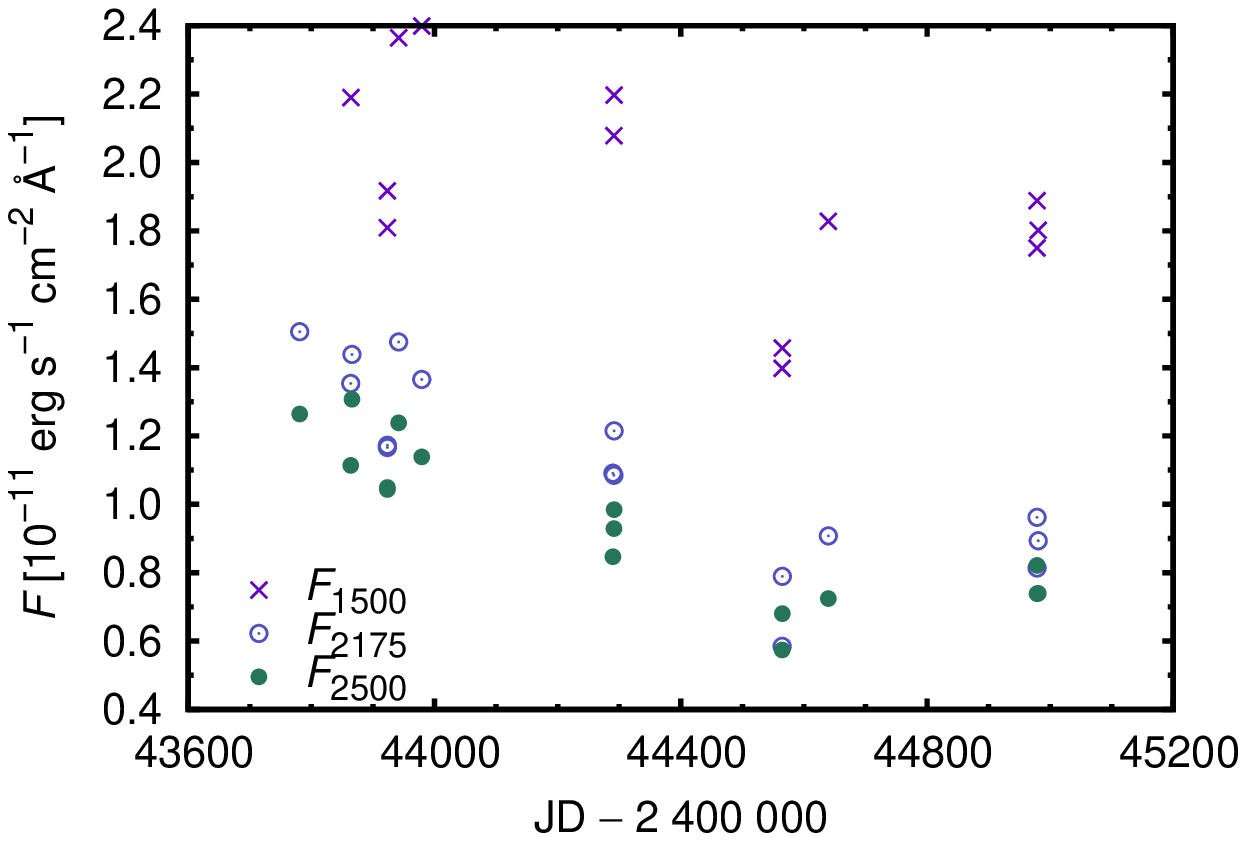}}
\resizebox{0.9\hsize}{!}{\includegraphics{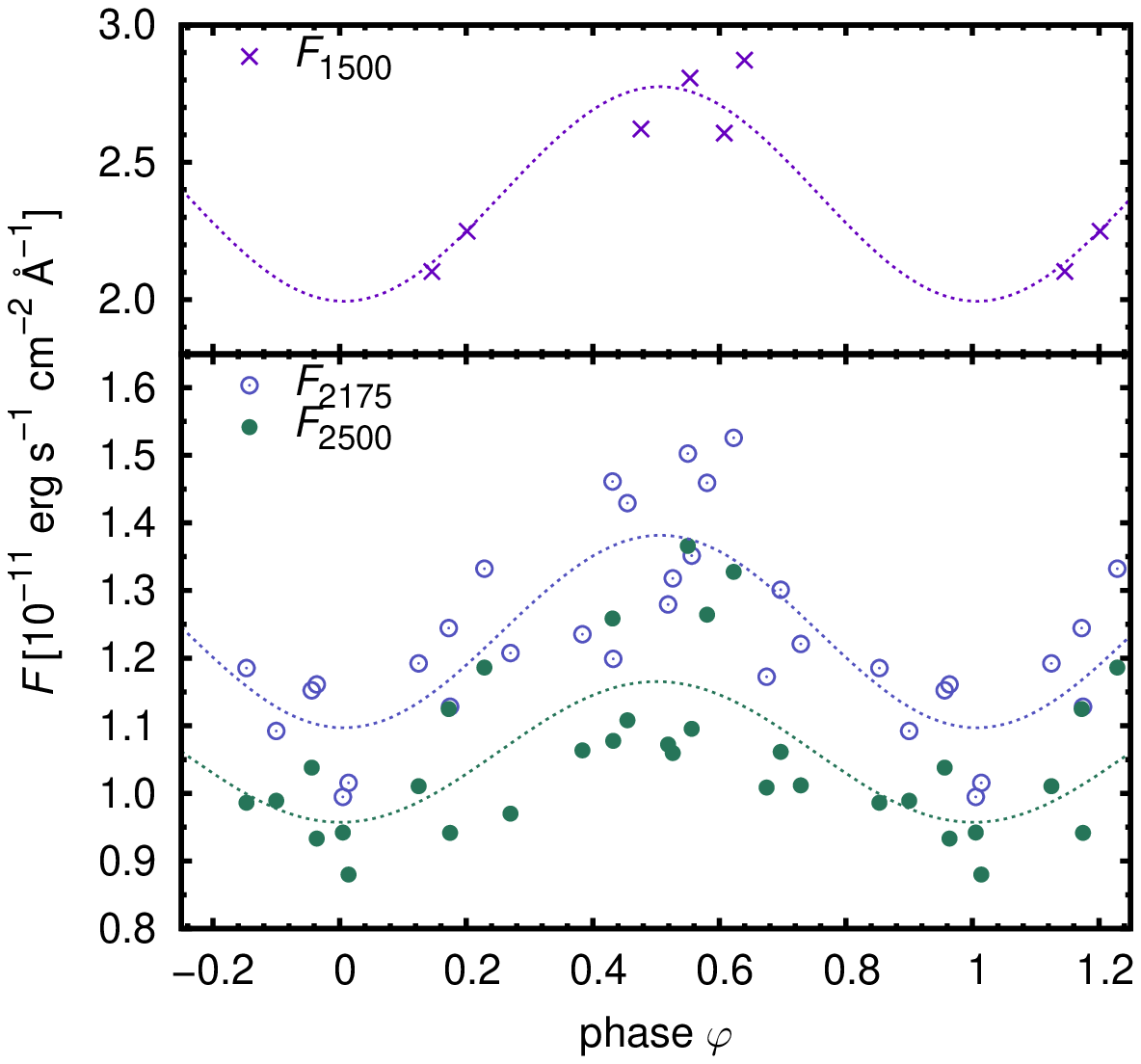}}
\caption{Variation of \hvezda\ broad-band fluxes. {\em Upper panel}: Long-term variations
during MJD 43\,800 -- 45\,000. {\em Bottom panel}: Periodic variations during
MJD 49\,400 -- 49\,800 plotted with ephemeris JD=$2\,449\,701.18\pm1.194E$.}
\label{hd50138prom}
\end{figure}

The IUE observations of \hvezda\ cover two epochs with MJD of about 43\,800 --
45\,000 and 48\,600 -- 49\,800. The stellar flux systematically decreased during
MJD 43\,800 -- 45\,000 in all bands studied here, but the short-term variations
are also significant during that epoch (see Fig.~\ref{hd50138prom}). These short
term variations are apparent in the Hipparcos photometry \citep{ESA} and
possibly also in the visual photometry \citep{halbedel}. The flux reached
original level during MJD 48\,600 -- 49\,800. The long-term variations could be
probably explained by increasing obscuration by the matter expelled during
outburst \citep{hutse}, which was already dispersed in the second epoch.

We used Period04 \citep{period} to analyse the $F_{2175}$ and $F_{2500}$ data in
the epoch MJD 49\,400 -- 49\,800 yielding the detection of the period
$1.194\pm0.006\,\text{d}$. A similar frequency is present also in the $F_{1500}$
data albeit with much larger uncertainty. All variations can be nicely fitted by
simple sinusoidals (see Fig.~\ref{hd50138prom}). We have checked the data from
MJD 43\,800 -- 45\,000 and data from Hipparcos (taken in MJD 47\,900 -- 49\,100)
and we have not found any reasonable periodicity.

The total emission and absorption line fluxes also show strong variations for
MJD 43\,800 -- 45\,000 (see Fig.~\ref{hd50138emabs}). The inspection of the
total line absorption in the epoch MJD 49\,400 -- 49\,800 also shows
possible periodicity, but with relatively large scatter. However, the total
emission line fluxes show steady increase for MJD 49\,400 -- 49\,800 when the
periodical photometric variations are detected.

The stochastic variations during MJD 43\,800 -- 45\,000 may be the consequence
of the outburst that appeared in 1978--1979 \citep{hutse}. \citet{borecferdapul}
attributed the detected photospheric variations to pulsations. If this model is
correct, then the fact that the pulsations are detectable only during the epoch
when the star does not show stochastic variations likely means that either the
pulsations are triggered by stochastic variations or are hidden in these
variations. \citet{krauhac} suggested that pulsations in a blue supergiant
55~Cygni can lead to phases of enhanced mass loss. In \hvezda, this could be
possibly connected to the increase of the total emission line flux
(Fig.~\ref{hd50138emabs}). The pulsational period is close to that found in
main-sequence stars with similar spectral type \citep{diag,nein} indicating that
\hvezda\ is still main-sequence star \citep[in agreement with][]{borecferdahp}.

For MJD>49\,000 the observations possibly evince the phase variability
of \tuv\ from about $11\,800\pm200\,$K to about
$11\,400\pm200\,$K. The region close to the
Ly$\alpha$ line shows a relatively broad (at about 1150 -- 1250\,\AA) flux
minimum, which is connected with low \tuv.

\subsection{HD 87643}

\renewcommand\hvezda{HD 87643}

\begin{figure}
\centering
\resizebox{0.9\hsize}{!}{\includegraphics{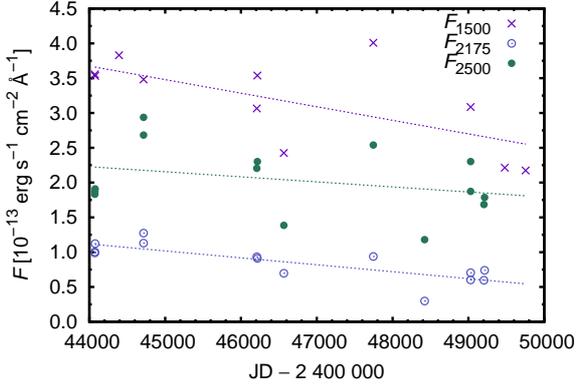}}
\caption{Variation of \hvezda\ broad-band fluxes with time. Lines denote linear fit to
individual broad-band fluxes.}
\label{hd87643prom}
\end{figure}

\begin{figure}
\centering
\resizebox{0.9\hsize}{!}{\includegraphics{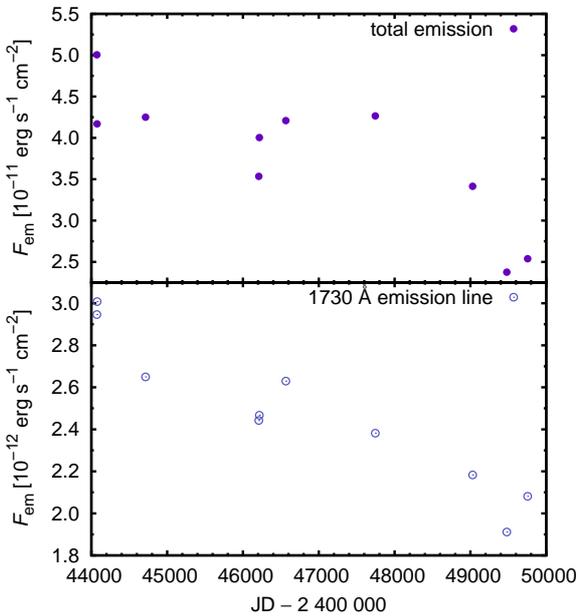}}
\caption{Variation of \hvezda\ total emission line flux ({\em upper panel}) and
the flux in the emission line at 1730\,\AA\ ({\em lower panel}).}
\label{hd87643em}
\end{figure}

The IUE observations of \hvezda\ in Fig.~\ref{hd87643prom} covering the period
of MJD 44\,000 -- 50\,000 show an overall decrease of the flux with possibly
additional stochastic variations. We fitted the flux variations with linear
function
$F_\lambda(\text{MJD})=a_\lambda(\text{MJD}-\overline{\text{MJD}})+b_\lambda$,
where $\overline{\text{MJD}}$ is the mean MJD for each observational set. The
wavelength dependence of $a_\lambda$ and $b_\lambda$ can be explained by the
variations of either dust absorption or temperature. The decrease of the
broad-band fluxes is accompanied by the decrease of the total and $1730$\,\AA\
emission line fluxes (see Fig.~\ref{hd87643em}). A similar decrease appears in
$1590$\,\AA\ emission line flux. The total line absorption varies between MJD 44\,000 and
MJD 49\,800 with the maximum roughly in the middle of the time interval and with
the minima at the beginning and the end of the observational time period 
(Table~\ref{slephd87643}).

The fitting of the flux reveals a possible increase of the \tuv\ during the
studied period from about $10\,900\pm200\,$K to $11\,700\pm800\,$K.

\subsection{HD 94878}

\renewcommand\hvezda{HD 94878}

Despite the binary nature of the object, \hvezda\ does not show any clear
periodicity of UV fluxes. Even the phase diagram based on the ephemeris of
\citet{brezina} does not show any clear trend in any colour. This is probably
connected with low amount of data and with the fact than the visual light
variations show a large scatter \citep{gosetvudoli,dalsivani} possibly as a
result of the intrinsic variability of a component. This component likely
dominates in UV making the binary lightcurve undetectable. The broad-band
photometry does not show any long-term variability in the studied time interval.

Despite the lack of clear flux variability, the absorption lines show
variability which may be orbitally modulated. The \ion{C}{ii} 1335\,\AA\
\citep{bragos} line absorption is variable (see Table~\ref{slephd94878})
with possible minimum
around the phase 0.3 of \citet{brezina}. On the other hand, the \ion{C}{iv}
absorption line at 1548\,\AA\ does not evince orbitally modulated variability. The
total emission and absorption fluxes suggest variability
(Table~\ref{slephd94878}).
The broad-band flux $F_{2175}$ is relatively low compared to other fluxes, which
is caused by the dust absorption.

\subsection{HD 100546}

\renewcommand\hvezda{HD 100546}

The UV spectrum of \hvezda\ does not show any clear flux variability. A small
dispersion of UV flux of about 6\,\% is consistent with small amplitude of
optical variability \citep{vandende}. There may be increase of the flux in the
emission line at 1779\,\AA\ and the total emission line flux from MJD=48\,000 to
50\,000 (Table~\ref{slephd100546}). There may be
also some variability in 1657\,\AA\ \ion{Fe}{ii} absorption line. Other
absorption lines as \ion{C}{ii} 1335\,\AA\ and the total
line absorption do not show any variability (Table~\ref{slephd100546}).

\subsection{HD 169515}

\renewcommand\hvezda{HD 169515}

The stellar spectral
energy distribution of \hvezda\ evince strong influence of the dust absorption on
$F_{2175}$ fluxes, which are by order of magnitude lower than $F_{2500}$
fluxes. We have not detected any clear long term flux variability of this star,
however, we are able to detect the 11.1~d orbital period \citep{krajan} from the
1500\,\AA\ and 2500\,\AA\ flux data.


\begin{figure}
\centering
\resizebox{0.9\hsize}{!}{\includegraphics{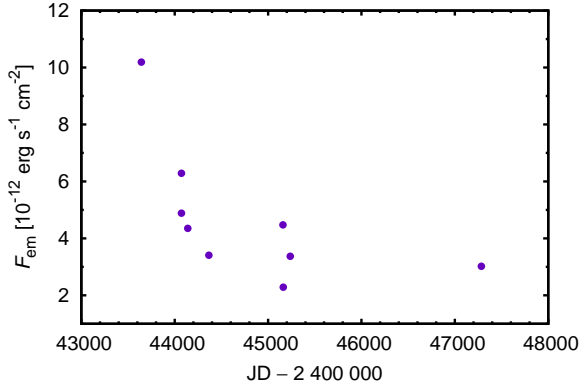}}
\caption{Variation of \hvezda\ total emission line flux.}
\label{hd169515em}
\end{figure}

The total emission and absorption fluxes do not show any strong phase
dependence, but there may be long-term variations of the total fluxes. The
decrease of the total emission line flux during the observational period is
shown in Fig.~\ref{hd169515em}. The total line absorption displays slightly
downward trend during the observational period. Individual
lines show phase dependence. The \ion{Fe}{ii} absorption line at
1658\,\AA\ \citep{zahada2} is the weakest during visual eclipses \citep[using
ephemeris of][]{krajan}. We have not detected any phase variability of
\ion{C}{iv} emission lines at $1550\,$\AA\ \citep{zahada2}, but all studied
lines evince variability, which means \ion{Si}{iv} 1402\,\AA\ emission
line, \ion{C}{iv} $1550\,$\AA\ emission line, and \ion{Fe}{ii} 1658\,\AA\
absorption line (Table~\ref{slephd169515}).

\section{General relations between individual fluxes}

\begin{figure}
\centering
\resizebox{0.9\hsize}{!}{\includegraphics{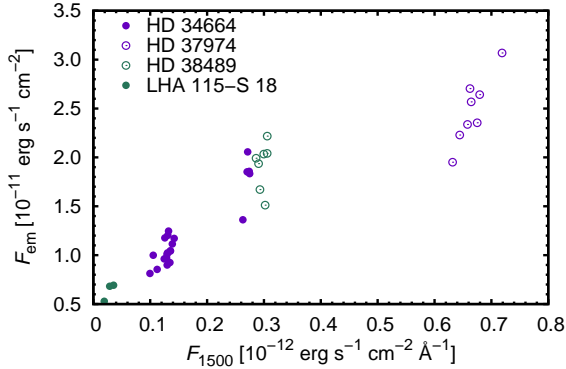}}
\caption{Relation between the far-UV broad band fluxes and the total emission
line fluxes for the stars from the Magellanic Clouds.}
\label{emisekubroadzMracen1}
\end{figure}

Fig.~\ref{emisekubroadzMracen1} shows that there is correlation between the
far-UV broad band fluxes and the total emission line fluxes for the stars from
the Magellanic Clouds. The emission line flux depends on the amount of ionizing
radiation and therefore on the stellar effective temperature and radius.
However, these parameters can not be the cause of the relationship in
Fig.~\ref{emisekubroadzMracen1}, because in limit of zero broad-band flux
$F_{1500}$ the emission line flux $F_\text{em}$ does not approach zero.
Consequently, it is more likely that the relationship results from geometrical
reasons, i.e., from varying inclination or radius of the disk. The origin of
continuum and some emission lines is different. Therefore, the relationship in
Fig.~\ref{emisekubroadzMracen1} could be possibly explained assuming that with
larger disk column density (or with higher disk inclination, i.e.,
for disks seen edge-on) the flux gets more absorbed and therefore $F_{1500}$
becomes lower. However, some parts of the envelope are visible even for 
high inclinations or large disk radii, consequently, $F_\text{em}$ is nonzero
for very low $F_{1500}$. Because the amount of ionizing radiation depends mainly
on the effective temperature, a relatively tight correlation of $F_{1500}$ and
$F_\text{em}$ fluxes indicates that the supergiant B[e] stars in the Magellanic
Clouds come from a relatively narrow range of the stellar parameters.

\begin{figure}
\centering
\resizebox{0.9\hsize}{!}{\includegraphics{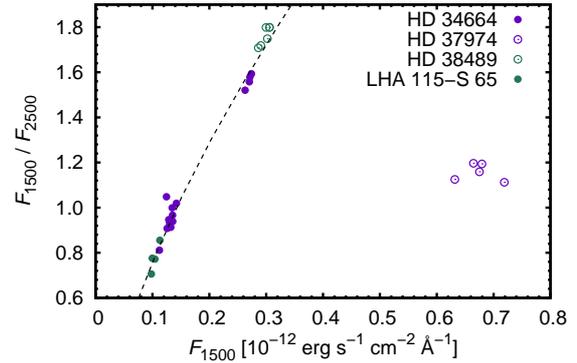}}
\caption{Relation between the far-UV broad band fluxes and the ratio of far-UV
and near-UV broad band fluxes for the stars from the Magellanic Clouds. Dashed
line denotes $F_{1500}\sim T_\text{eff}^4 F_{1500}/F_{2500}$ relation derived
from \mbox{ATLAS} model atmospheres.}
\label{emisekubroadzMracen2}
\end{figure}

With exception of HD~37974, the ratio of far-UV and near-UV broad band fluxes
for stars from the Magellanic Clouds depends on the far-UV flux (see
Fig.~\ref{emisekubroadzMracen2}). We have not included LHA 115-S~18 in the plot,
because the star does not show a strong continuum. The variations in
Fig.~\ref{emisekubroadzMracen2} show a tight linear trend for all stars except
HD~37974 despite complicated flux variations in individual stars. The relation
can be fitted using model atmospheres emergent fluxes $f_{1500}$ and $f_{2500}$
\citep{kurat} assuming constant luminosity of B[e] stars and variable UV
temperature and radius $R_\text{UV}$, as $L\sim R_\text{UV}^2T_\text{UV}^4$ or
$R_\text{UV}^2\sim T_\text{UV}^{-4}$ for constant $L$. From this follow
relationships $f_{1500}/T_\text{UV}^4\sim f_{1500}/f_{2500}$ or $f_{1500}\sim
T_\text{UV}^4 f_{1500}/f_{2500}$ for the model atmosphere fluxes, or
(multiplying by $R_\text{UV}^2$)  $F_{1500}\sim F_{1500}/F_{2500}$. The flux
$f_{1500}$ in Fig.~\ref{emisekubroadzMracen2} was linearly scaled to fit the
observed relation and $T_\text{UV}$ is the effective temperature $T_\text{eff}$
corresponding to model atmosphere flux. These results show that a large group of
B[e] supergiants has the same luminosity of about
$(1.9\pm0.4)\times10^5\,L_\odot$ from the model atmospheres, assuming the LMC
distance 50\,kpc \citep{deg}, and correcting for mean extinction derived here.
This can provide the possibility to reveal the origin of B[e] supergiant stars.
Moreover, with a proper calibration, the relationship in
Fig.~\ref{emisekubroadzMracen2} could be used for distance estimation. 

\begin{figure}
\centering
\resizebox{0.9\hsize}{!}{\includegraphics{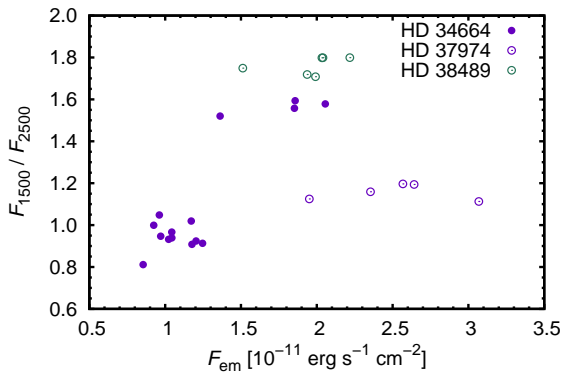}}
\caption{Relation between the total emission line flux and the ratio of far-UV
and near-UV broad band fluxes for the stars from the Magellanic Clouds.}
\label{emisekubroadzMracen3}
\end{figure}

Fig.~\ref{emisekubroadzMracen3} shows that there is not a clear correlation
between the emission line flux and the ratio of $F_{1500}/F_{2500}$, which is a
temperature indicator. The lack of correlation supports a general picture that
the continuum flux observed in the UV region does not originate in the stellar
atmosphere.

\begin{figure}
\centering
\resizebox{0.9\hsize}{!}{\includegraphics{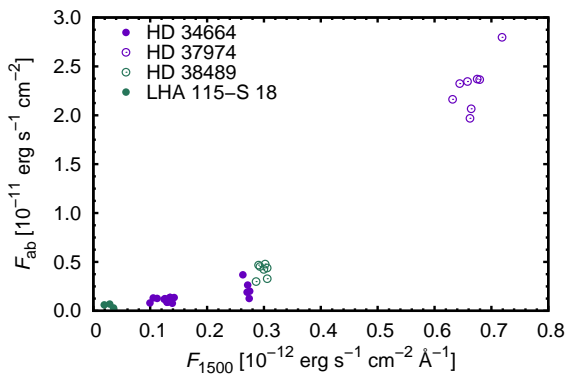}}
\caption{Relation between the broad-band flux at $1500$\,\AA\ and the total
line absorption for the stars from the Magellanic Clouds.}
\label{emisekubroadzMracen4}
\end{figure}

The total line absorption is strongly correlated with the $F_{1500}$ flux for
the stars from the Magellanic Clouds (see Fig.~\ref{emisekubroadzMracen4}). The
$F_{1500}$ flux increases with increasing temperature of the circumstellar
envelope, consequently, from the nonlinearity of the relationship in 
Fig.~\ref{emisekubroadzMracen4} follows that the absorption lines become
relatively stronger for higher temperature of the envelope.

%
%

\section{Discussion and conclusions}

We studied the UV variability of B[e] stars in individual lines and in
broad-band fluxes. From the observed flux distribution, we calculated the
broad-band variations using artificial photometric filters centered on selected
wavelengths.
We evaluated the total emission and absorption line fluxes and
studied their variability, which originates in the stellar envelope, disk, and
the star alone. The observed spectral energy distributions were fitted with
atmosphere flux attenuated by the dust absorption.

The UV domain provides wealth of information about the variability of B[e]
stars. The multi-colour photometry centered ideally in the far-UV and near-UV
regions is able to detect long-term variability (in LHA~115-S~18, HD~37974,
HD~45677, HD~50138, and HD~87643) and the S~Doradus type variability (in the
case of HD~34664). The detected eclipses in HD~169515 may help to constraint the
nature of companions. Some stars also show irregular short-term variability
(HD~94878). In HD~50138, we also detected pulsations, which are relatively rare
among B[e] stars. The derived pulsation period in HD~50138 is $1.194\,\text{d}$.
This is the first unambiguous detection of the pulsation and measurement of its
period in this star. Some B[e] stars do not show any strong flux variability
(LHA~115-S~65, HD~38489, and HD~100546).

The line variability is typically connected with the flux variability. In
HD~34664 the decrease of the flux (and the estimated temperature) is correlated
with the decrease of the line fluxes. A similar correlation is also found in
HD~37974, HD~45677, and HD~87643. LHA~120-S~12 shows wind line profile
variability that corresponds to the wind bistability jump. In HD~169515 we have
found variability of absorption lines during the eclipses.

We found striking differences in the ratios of the total emission line flux to
the total line absorption among studied stars perhaps as a consequence of
envelope properties. The stars with high total emission line flux are LHA 115-S
18, LHA 115-S 65, HD 34664, HD 38489, HD 45677, HD 50138, HD 87643, and HD
169515. This group of stars shows the ratio of the total emission line flux to
the total line absorption $F_\text{em}/F_\text{ab}$ in the range 1.1--23.5.
The predominance of emission line flux is typical for these stars. On the
contrary, the stars LHA 120-S 12, HD 37974, HD 94878, and HD 100546 show low
ratio of the total emission line flux to the total line absorption and the
spectra evince more absorption lines than emission lines. We specify the ratio
$F_\text{em}/F_\text{ab}$ in the range 0.46--1.37 for these stars.

In many B[e] stars, the temperature obtained from the fitting of UV flux
distribution is significantly lower than 20~kK. This is too low to cause a
significant H$\alpha$ emission. This indicates that many B[e] stars are
enshrouded by optically thick envelope, and a significant part of UV radiation
does not originate on the surface of B[e] stars, but in their disk. The stellar
radiation is heavily obscured in such stars, and the temperature derived from
fitting corresponds to the temperature of the circumstellar environment.
This conclusion is further supported by the relation between the broad-band
fluxes and total emission line flux.
Possibly, the absorbing material appears also above the disk.

The fitted amount of the dust absorption derived from the 2175\,\AA\ opacity
bump does not vary with time in majority of stars. This indicates the
interstellar origin of this feature in many stars.

The fluxes from the Magellanic Cloud stars show that many B[e] stars
evince the same luminosity of about $1.9\times10^5\,L_\odot$ and similar
effective temperature. This indicates that the long-term broad-band variability
of B[e] stars is most likely caused by the changes in the envelopes of these
stars. Moreover, this puts strong constraints on the models of evolution of
these stars and possibly establishes distance indicator.



\section*{Acknowledgements}
This research was supported by GA\,\v{C}R 16-01116S.

\onecolumn

\appendix

\section{Basic properties of studied B[e] stars, lists of the IUE observations
used in the paper and derived broad-band and line fluxes, and supplementary
figures}
\label{slepak}

Here we summarize the basic properties of studied B[e] stars.

\renewcommand\hvezda{LHA 115-S 18}
\subparagraph{\hvezda}
The forbidden emission lines in the spectra of
SMC star
\hvezda\ (AzV 154) were detected
by \cite{azzo}. Besides the prominent variability of \ion{He}{ii} 4686\,\AA\
emission line \citep{sanda} and many UV lines \citep{shorenemapravdu}, the star
also shows optical variability on different time scales
\citep{vang,genster,uradnik}. A two-component model for the \hvezda\ outflow
consisting of fast radiatively driven wind and slow outflowing disk was proposed
by \citet{zikis18}. The star shows a strong infrared excess due to the dust
\citep{kasar}. The optical spectra of the star shows Raman-scattered lines
\citep{belana}. \citet{uradnik} proposed a binary nature of the object based on
the detection of X-ray emission \citep{tonda}.

\renewcommand\hvezda{LHA 115-S 65}
\subparagraph{\hvezda}
In the case of SMC B[e] supergiant LHA 115-S 65 (CPD-75$^\circ$ 116, RMC 50, Sk
193) any variability from IUE spectra has not been found \citep{shorsan2}. The
profiles of emission lines of this star are consistent with their origin in
Keplerian disk \citep{krauss65,arets65}.

\renewcommand\hvezda{LHA 120-S 12}
\subparagraph{\hvezda}
The LMC blue supergiant star \hvezda\ (Sk -67 23) shows H$\alpha$ emission and
dust envelope \citep{stahnul}. The detected polarization signature
\citep{moreplavec} points to the presence of non-spherical envelope. Measurement
of $^{13}$C relative abundance in the infrared spectra of this star provided
support for the evolved nature of the object \citep{izotopy}. 

\renewcommand\hvezda{HD 34664}
\subparagraph{\hvezda}
The B[e] supergiant HD~34664 (LHA 120-S 22, MWC 105) is a member of NGC~1871
association in the LMC and shows forbidden emission lines \citep{skorobach}. The
IUE spectra of this star were studied by \citet{bensa}. The polarimetry and
spectropolarimetry revealed the existence of a disk around this star
\citep{moreplavec,schula}. The nebula around the supergiant, which is visible in
the optical, H$\alpha$, and infrared wavelengths, reflects the H$\alpha$ line of
the star \citep{chuchen}. The structure of the nebula around the star is
especially complex in the IRAC observations of SPITZER.

\renewcommand\hvezda{HD 37974}
\subparagraph{\hvezda}
The IUE spectra of the LMC B[e] supergiant \hvezda\ (LHA 120-S 127) show very
strong resonance lines \citep{shorsan2}. This star, which is viewed pole-on,
displays signatures of the dusty disk in infrared and numerous \ion{Fe}{ii}
emission lines in near-UV \citep{dalsiziki,arets65}. The total mass of the dusty
disk is estimated to be $3\times10^{-3}\,M_\odot$ from SPITZER \citep{kabus}.

\renewcommand\hvezda{HD 38489}
\subparagraph{\hvezda}
The detailed study of IUE spectra \citep{shorsan} of LMC supergiant \hvezda\
(LHA\,120-S\,134) revealed presence of the stellar wind from P~Cygni lines
\citep{shorsan}. \citet{stahnul} detected the infrared dust emission in the
photometry of this star. Infrared observations also show a nebula around the
star \citep{kasar}. Some emission lines display double peaked shapes indicating
their origin in rotating disk \citep{arets65}. The infrared spectra of the star
show CO emission \citep{mcjirka,oxahlava}.

\renewcommand\hvezda{HD 45677}
\subparagraph{\hvezda}
Although HD 45677 (FS CMa) is classified as a B[e] star, its luminosity class is
not in agreement with the supergiant characteristics \citep{cid}. The UV energy
distribution shows signatures of dust absorption in HD 45677
\citep{odpadnici,sitnaodpad}. The UV spectropolarimetry of this star obtained
during space shuttle mission \citep{uvpolar} revealed a bipolar reflection
nebula consistent with dusty disk. \citet{stupne} interpreted available
observations concluding that HD 45677 is a Herbig Be star in the phase of
circumstellar material accretion. This was questioned by \citet{mirotrida},
partly due to the missing pre-stellar objects in the neighborhood. The infrared
properties of the star are also different from pre-main sequence objects
\citep{dalsilee}. \citet{sitovygraf} interpreted the observed UV, optical, and
infrared variability as a result of variable dust extinction.
The UV flux variations show increase from MJD of about 44\,000 to
48\,000 discussed by \citet{sitovygraf}. This corresponds to the recovery from a
deep visual minima, that occurred around 1980 \citep{zimajda,patel} and is
accompanied by the decrease of thermal emission in infrared
\citep{sitovygraf}. These variations are interpreted as a result of decreasing
obscuration by circumstellar dust, which was possibly ejected after an explosive
event around 1950 \citep{sitovygraf,zimajda}.

\renewcommand\hvezda{HD 50138}
\subparagraph{\hvezda}
The UV observations of \hvezda\ (V743 Mon) star were described by
\citet{simes}. The disk of the star was resolved using spectropolarimetry
\citep{bjork} and interferometry \citep{borecferda,eller}. The star shows
complex line profile variations that were attributed to the pulsations with
period significantly shorter than the rotational period of about 3.6\,d
\citep{borecferdapul}. The UV variability of this star was reported by
\citet{odpadnici} based on the observations of the ANS satellite. \citet{hutse}
provide evidence of outburst in 1978--1979 from IUE line data. The photometric
monitoring has not revealed a significant optical variability
\citep{moreponiku,halbedel}.

\renewcommand\hvezda{HD 87643}
\subparagraph{\hvezda}
The southern emission line star \hvezda\ (V640 Car) is a member of a wide binary
\citep{milour}, resides close to the \ion{H}{ii} region RCW 47 \citep{cramp},
and is accompanied by a reflection nebula clearly visible in the optical images
\citep{surdejsurdej}. \hvezda\ is one of the stars whose observations led to the
definition of the B[e] phenomenon \citep{swing,alles}. IUE observations of
\hvezda\ show strong lines of \ion{Fe}{ii} \citep{dfp}. The spectropolarimetric
observations \citep{oud} revealed the existence of the disk around \hvezda. The
star shows irregular light variations \citep{gr}.

\renewcommand\hvezda{HD 94878}
\subparagraph{\hvezda}
The eclipsing binary of $\beta$~Lyrae type \hvezda\ (GG Car, \citealt{brezina})
shows infrared excess \citep{alles}. The UV spectrum displays numerous lines of
ions with low ionization potential \citep{bragos,bragoss}. The polarimetric
properties of the star indicate the presence of multicomponent envelope
\citep{hnedin}. The H$\alpha$ spectropolarimetry reveals the presence of the
disk \citep{perak}, which rotates with Keplerian velocity \citep{dzidzi}. The
new data from GAIA DR1 \citep{gaia}, which give the distance $r>2\,$kpc, agree
with other studies \citep[e.g.,][]{brezina} and confirm luminous nature of the
source.

\renewcommand\hvezda{HD 100546}
\subparagraph{\hvezda}
\hvezda\ (KR Mus) is classified as a Herbig Ae/Be star \citep{hucaj}. The star
hosts a protoplanetary disk \citep{gar}.

\renewcommand\hvezda{HD 169515}
\subparagraph{\hvezda}
The double-lined eclipsing binary \hvezda\ (RY Sct) consists of mass losing
O9.7~Ibep supergiant accompanied by massive companion
\citep{ankum,koren,djurak}. The binary interaction provides a natural
explanation of the H$\alpha$ emission line and other spectral features connected
with the B[e] phenomenon. The observed ring structure of the
circumstellar medium indicates possible discrete mass ejection phases
\citep{mensik,agent}.
The IUE spectrum of \hvezda\ was extensively studied by \citet{zahada1,zahada2},
who provide also identification of many UV lines.

\renewcommand\hvezda{LHA 115-S 18}
\begin{figure}
\centering
\resizebox{0.45\hsize}{!}{\includegraphics{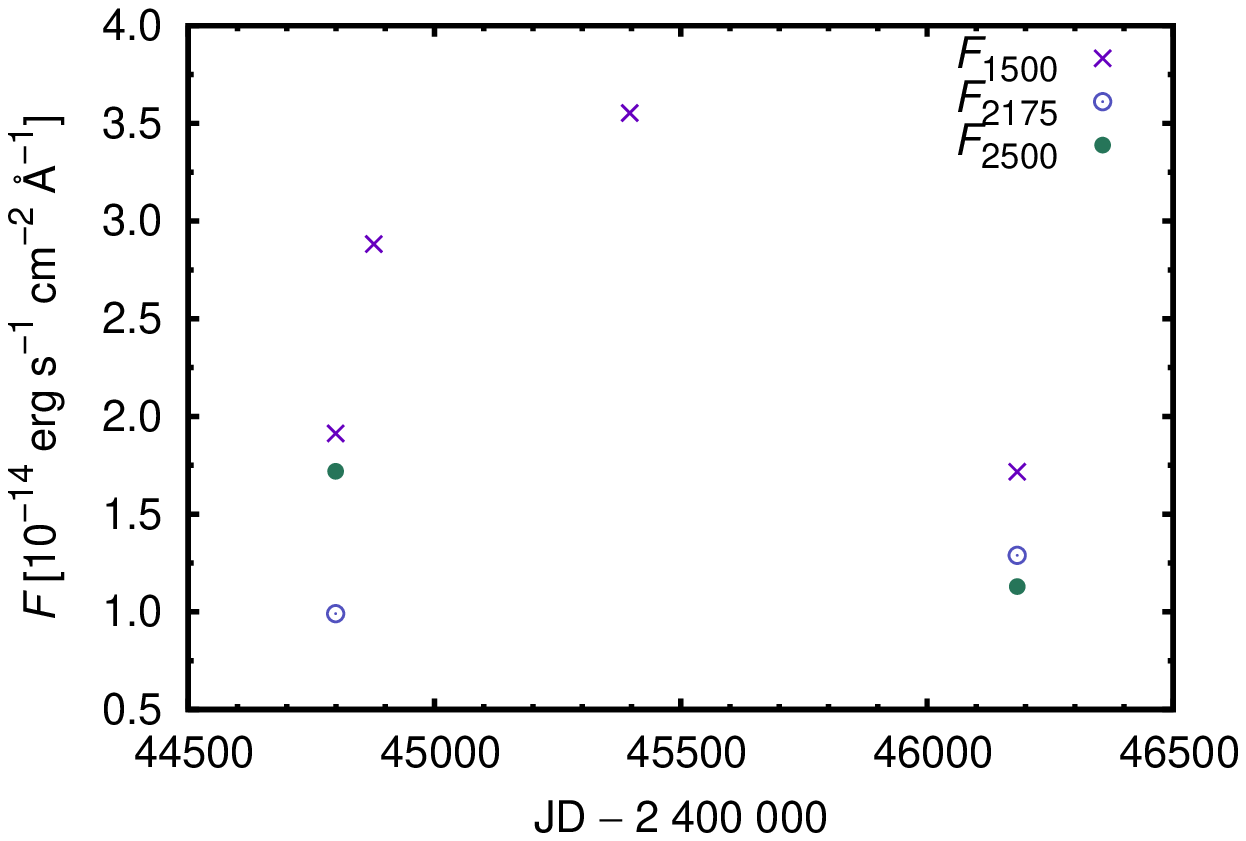}}
\caption{Variation of \hvezda\ broad-band fluxes with time.}
\label{lha115s18}
\end{figure}

\renewcommand\hvezda{HD 37974}
\begin{figure}
\centering
\resizebox{0.45\hsize}{!}{\includegraphics{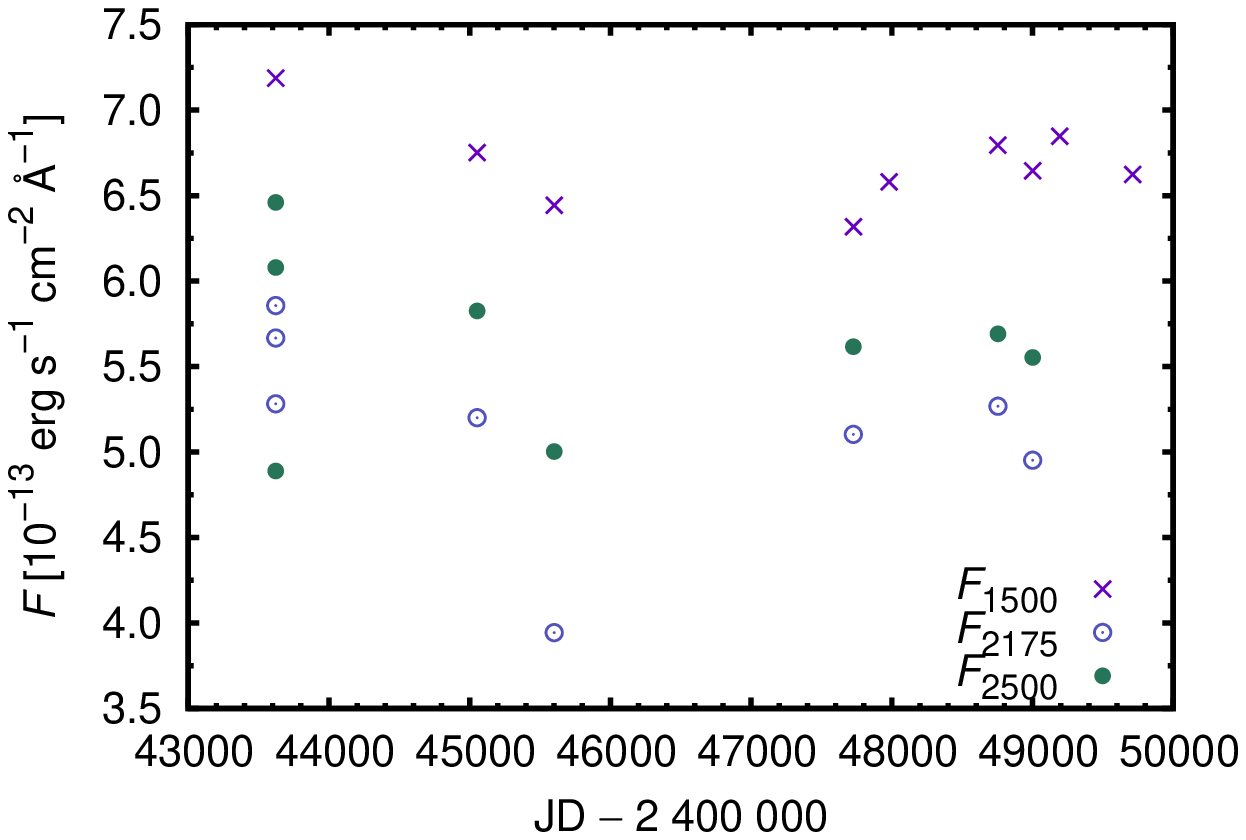}}
\resizebox{0.45\hsize}{!}{\includegraphics{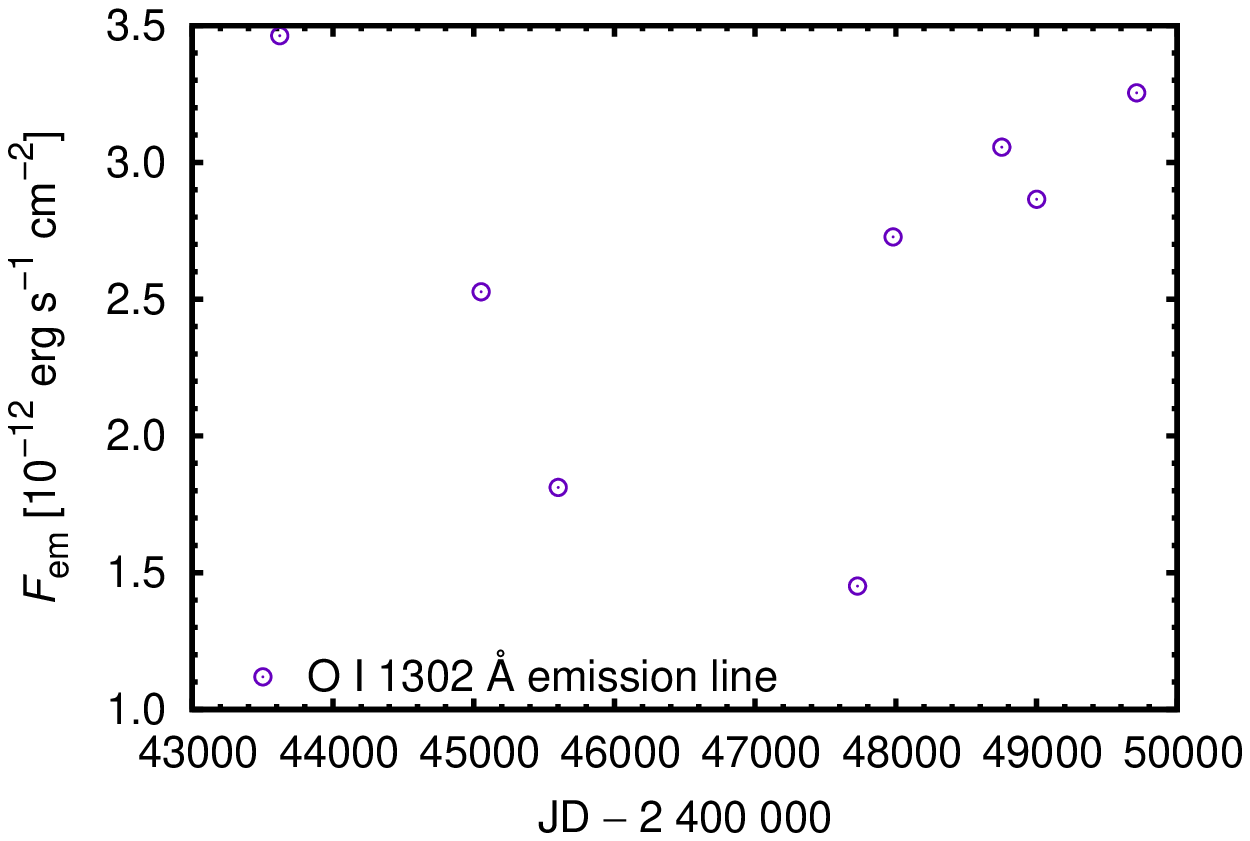}}
\caption{Variation of \hvezda\ with time. {\em Left panel}: Broad-band fluxes.
{\em Right panel} \ion{O}{i} 1302\,\AA\ emission line flux.}
\label{hd37974prom}
\end{figure}

\renewcommand\hvezda{HD 50138}
\begin{figure}
\centering
\resizebox{0.45\hsize}{!}{\includegraphics{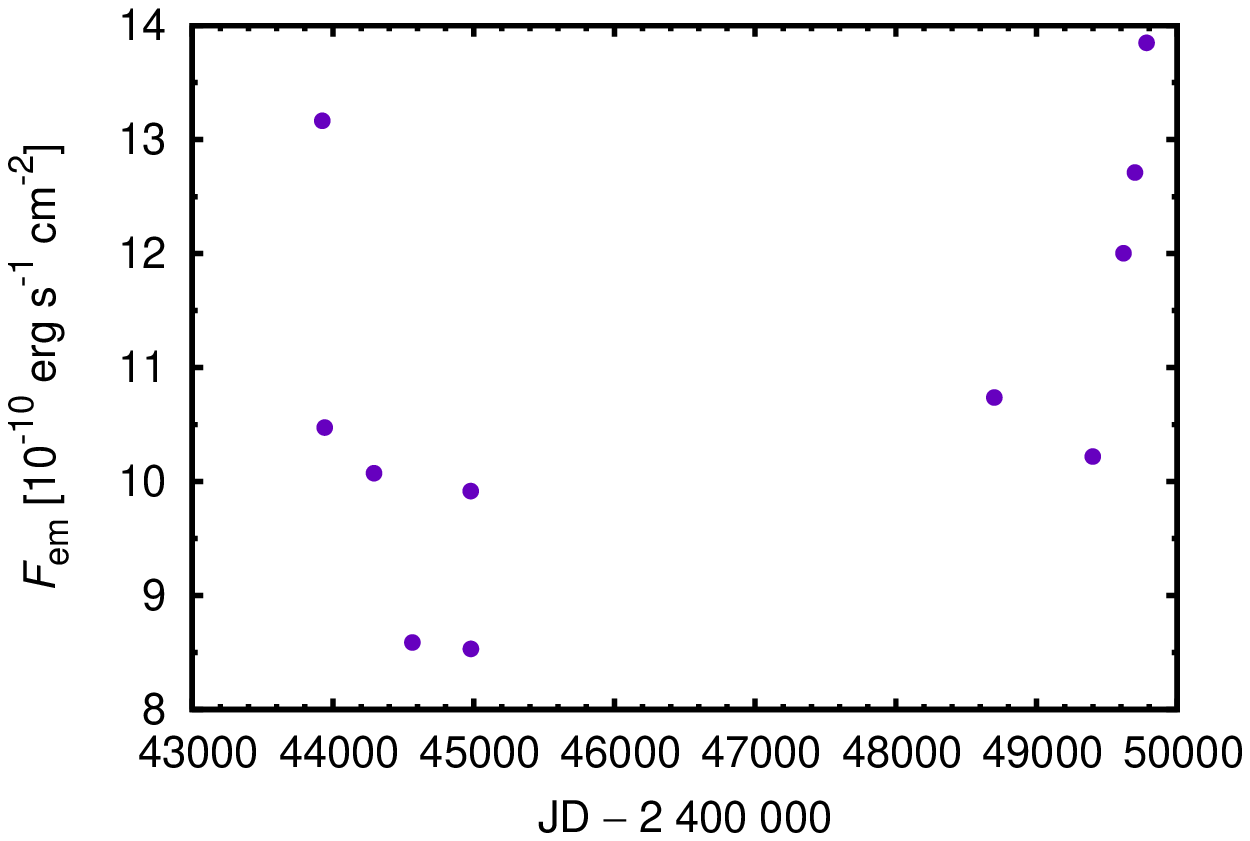}}
\resizebox{0.45\hsize}{!}{\includegraphics{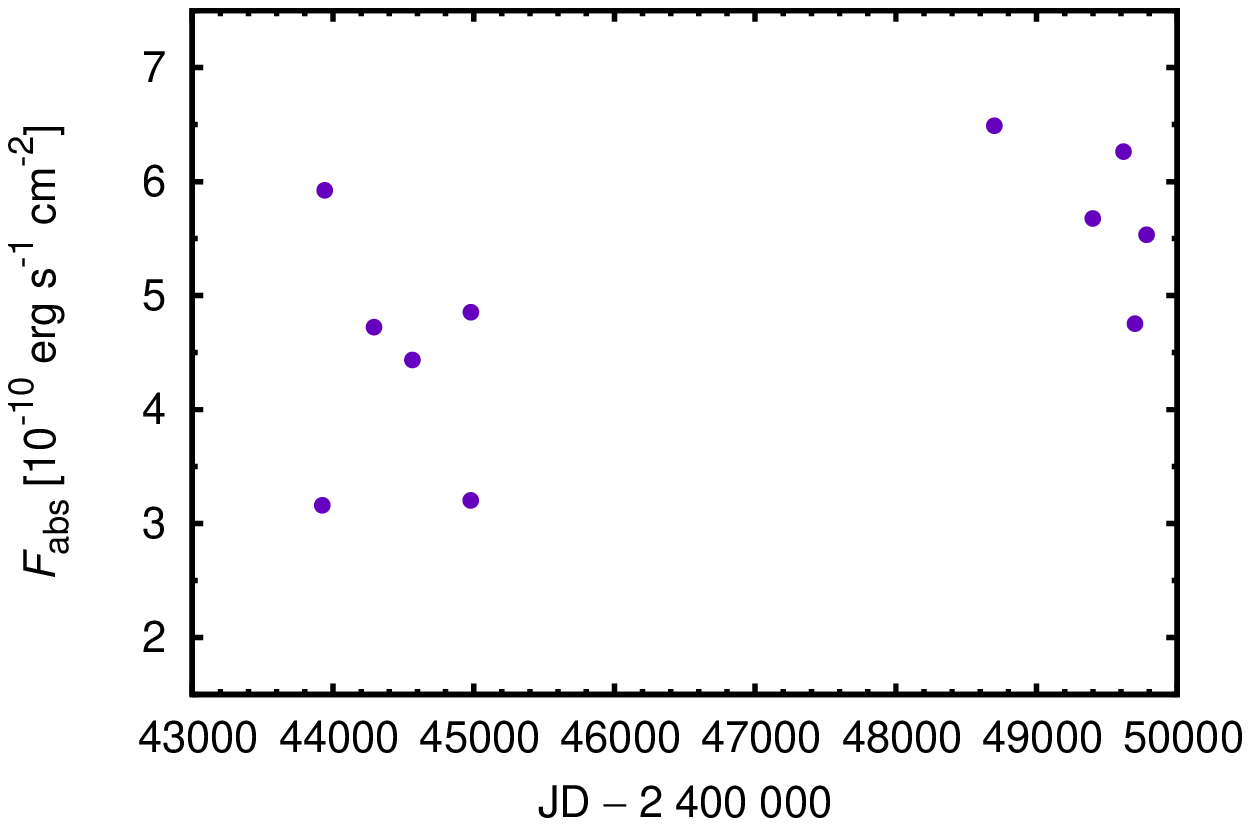}}
\caption{Variation of \hvezda\ total emission line fluxes ({\em left panel})
and total line absorption ({\em right panel}) with time.}
\label{hd50138emabs}
\end{figure}

\begin{table}
\caption{List of the IUE observations and fluxes of LHA 115-S 18. The units of
broad-band flux $F_{1500}$ are $10^{-14}\,\ergsa$, the units of broad-band
fluxes $F_{2175}$ and $F_{2500}$ are 
$10^{-15}\,\ergsa$. The units of $F_\text{em}$ (for individual lines and the total
flux) and $F_\text{ab}$ (total) are $10^{-14}\,\ergs$.}
\label{sleps18}
\begin{center}
\begin{tabular}{cccccccccc}
\hline
Camera & Image & Julian date & $F_{1500}$ & $F_\text{em}$
&$F_\text{em}$ &$F_\text{em}$ &$F_\text{em}$ &$F_\text{em}$ &$F_\text{ab}$ \\
& & 2,400,000.5+& & \ion{N}{v} 1245\,\AA & \ion{O}{i} 1302\,\AA& \ion{C}{iv} 1550\,\AA
& \ion{He}{ii} 1640\,\AA & \multicolumn{2}{c}{total}\\
\hline
SWP & 14463 & 44798.59642 & 1.91 & 170 & 17 & 42 & 140 & 530 & 61\\
SWP & 15122 & 44876.45930 & 2.88 & 240 & 30 & 88 & 220 & 680 & 69\\
SWP & 19371 & 45395.84036 & 3.55 & 150 & 20 & 62 & 200 & 690 & 30\\
SWP & 25789 & 46182.47697 & 1.72\\
\hline
\multicolumn{3}{l}{mean uncertainty} & 0.7 & 30 & 6 & 18 & 15 & 160 & 13\\
\hline
\end{tabular}
\end{center}
\begin{center}
\begin{tabular}{ccccc}
\hline
Camera & Image & Julian date  & $F_{2175}$ & $F_{2500}$ \\
       &&   2,400,000.5+ \\
\hline
LWR & 11058 & 44798.64322 & 9.90 & 17.2\\
LWP & 05836 & 46182.55200 & 12.9 & 11.3\\
\hline 
\multicolumn{3}{l}{mean uncertainty} & 12 & 4\\
\hline
\end{tabular}
\end{center}
\end{table}

\begin{table}
\caption{List of the IUE observations and fluxes of LHA 115-S 65. The units of
broad-band flux $F_{1500}$ are $10^{-14}\,\ergsa$ and the units of broad-band
fluxes $F_{2175}$ and $F_{2500}$ are
$10^{-13}\,\ergsa$.}
\begin{center}
\begin{tabular}{cccc}
\hline
Camera & Image & Julian date  & $F_{1500}$\\
       &&   2,400,000.5+ \\
\hline
  SWP & 10990 &  44611.36670 &9.97\\
  SWP & 13503 &  44678.88659 &9.80\\
  SWP & 13504 &  44678.94138 &10.6\\
  SWP & 15121 &  44876.38032 &10.4\\
  SWP & 43360 &  48601.44294 &9.77\\
  SWP & 48257 &  49196.52540 &11.3\\
\hline
\multicolumn{3}{l}{mean uncertainty} & 1.4\\
\hline
\end{tabular}\qquad
\begin{tabular}{ccccc}
\hline
Camera & Image & Julian date & $F_{2175}$ & $F_{2500}$\\
       &&   2,400,000.5+ \\
\hline
  LWR & 09659 &  44611.38731 & 1.12 & 1.29\\
  LWR & 10146 &  44678.91174 & 1.24 & 1.39\\
  LWR & 11057 &  44798.54116 & 1.37 & 1.38\\
  LWR & 11638 &  44876.40194 & 1.26 & 1.35\\
  LWP & 21992 &  48601.73820 & 1.04 & 1.23\\
  LWP & 26031 &  49196.49965 & 1.10 & 1.32\\
\hline
\multicolumn{3}{l}{mean uncertainty} & 0.3 & 0.1\\
\hline
\end{tabular}
\end{center}
\end{table}

\begin{table}
\caption{Mean LHA 115-S 65 broad-band fluxes}
\label{iues65mag}
\begin{center}
\begin{tabular}{cccc}
\hline
Julian date & $F_{1500}$ & $F_{2175}$ & $F_{2500}$\\
2,400,000+ &
\multicolumn{3}{c}{[$10^{-13}\,\ergsa$]}\\
\hline
44730 & $1.02\pm0.04$ & $1.25\pm0.09$ & $1.35\pm0.04$\\
48900 & $1.05\pm0.13$ & $1.07\pm0.06$ & $1.28\pm0.08$\\
\hline
\end{tabular}
\end{center}

\end{table}

\begin{table}
\caption{List of the IUE observations and fluxes of LHA 120-S 12. The units of
broad-band flux $F_{1500}$ are $10^{-13}\,\ergsa$, units of broad-band fluxes $F_{2175}$ and $F_{2500}$ are
$10^{-14}\,\ergsa$, the units of $F_\text{ab}$ (\ion{Si}{iv} 1393\,\AA, \ion{C}{iv}
1548\,\AA) are $10^{-12}\,\ergs$, the units of $F_\text{em}$ (total) are
$10^{-12}\,\ergs$, and the units of $F_\text{ab}$ (total) are $10^{-11}\,\ergs$.}
\label{sleps12}
\begin{center}
\begin{tabular}{cccccccc}
\hline
Camera & Image & Julian date & $F_{1500}$ & $F_\text{ab}$ & $F_\text{ab}$ &
$F_\text{em}$ & $F_\text{ab}$\\
       &&   2,400,000.5+ & & \ion{Si}{iv} & \ion{C}{iv}&
       \multicolumn{2}{c}{total}\\
\hline
SWP & 14450 & 44796.80900 & 1.70 & 2.4 & 1.6 & 3.9 & 1.1\\
SWP & 36709 & 47727.77732 & 1.64 & 2.7 & 1.1 & 3.6 & 1.1\\
SWP & 38454 & 47978.52236 & 1.77 & 3.2 & 1.5 & 3.1 & 1.3\\
SWP & 38455 & 47978.58015 & 1.76 &     & 1.6 & 2.2 & 1.4\\
SWP & 44666 & 48756.93182 & 1.72 & 3.6 & 2.2 & 3.0 & 1.3\\
\hline
\multicolumn{3}{l}{mean uncertainty} & 0.1 & 0.1 & 0.2 & 0.3 & 0.1\\
\hline
\end{tabular}\qquad
\begin{tabular}{ccccc}
\hline
Camera & Image & Julian date & $F_{2175}$ & $F_{2500}$\\
       &&   2,400,000.5+\\
\hline
LWR & 11639 & 44876.51574 & 10.4 & 8.48\\
LWP & 17633&  47978.54468 & 11.0 & 9.59\\
\hline
\multicolumn{3}{l}{mean uncertainty} & 2.4 & 1.0\\
\hline \\ \\ \\
\end{tabular}
\end{center}
\end{table}

\begin{table}
\caption{List of the IUE observations of HD 34664. The units of broad-band
fluxes
$F_{1500}$, $F_{2175}$, and $F_{2500}$ are $10^{-13}\,\ergsa$, the units of
$F_\text{em}$ (\ion{O}{i} 1302\,\AA)
are $10^{-13}\,\ergs$, the units of $F_\text{em}$ (total) are $10^{-11}\,\ergs$,
and the units of $F_\text{ab}$ (total) are $10^{-12}\,\ergs$.}
\begin{center}
\begin{tabular}{ccccccc}
\hline
Camera & Image & Julian date & $F_{1500}$ & $F_\text{em}$ & $F_\text{em}$ &
$F_\text{ab}$\\
       &&   2,400,000.5+& &\ion{O}{i} 1302\,\AA& \multicolumn{2}{c}{total}\\
\hline
SWP & 07634 & 44250.42995 & 2.71 & 9.2 & 2.1 & 2.6\\
SWP & 07881 & 44276.32969 & 2.74 & 8.2 & 1.9 & 1.3\\
SWP & 13505 & 44678.98535 & 2.70 & 11.2& 1.9 & 1.9\\
SWP & 14449 & 44796.73076 & 2.75 & 6.5 & 1.8 & 2.0\\
SWP & 15114 & 44875.47450 & 2.63 & 7.4 & 1.4 & 3.7\\
SWP & 38609 & 47996.73495 & 1.42 & 7.6 & 1.2 & 1.4\\
SWP & 38667 & 48004.95577 & 1.35 & 9.8 & 1.0 & 1.4\\
SWP & 38668 & 48005.02173 & 1.39 & 8.4 & 1.1 & 0.8\\
SWP & 40926 & 48311.45845 & 1.32 & 8.9 & 1.2 & 0.9\\
SWP & 43125 & 48577.25947 & 1.30 & 6.5 & 1.0 & 1.1\\
SWP & 43349 & 48599.47516 & 1.26 & 9.1 & 1.2 & 1.2\\
SWP & 43350 & 48599.53042 & 1.25 & 8.2 & 1.0 & 1.2\\
SWP & 44160 & 48693.90032 & 1.31 & 7.0 & 1.2 & 0.9\\
SWP & 45240 & 48831.59244 & 1.35 & 6.2 & 0.9 & 1.4\\
SWP & 46592 & 48982.96380 & 1.29 & 5.4 & 1.0 & 1.2\\
SWP & 48202 & 49190.30118 & 0.995& 5.2 & 0.8 & 0.8\\
SWP & 48709 & 49253.41809 & 1.05 & 5.3 & 1.0 & 1.3\\
SWP & 50703 & 49478.58207 & 1.30 & 6.2 & 1.0 & 0.9\\
SWP & 50711 & 49479.43130 & 1.36 & 5.9 & 1.0 & 1.0\\
SWP & 53169 & 49710.28760 & 1.12 & 5.6 & 0.9 & 1.3\\
SWP & 54482 & 49829.01160 & 1.30 & 5.8 & 0.9 & 0.9\\
\hline
\multicolumn{3}{l}{mean uncertainty} & 0.1 & 1.2 & 0.1 & 0.2\\
\hline
\end{tabular}\qquad
\begin{tabular}{ccccc}
\hline
Camera & Image & Julian date & $F_{2175}$ & $F_{2500}$\\
       &&   2,400,000.5+& \\
\hline
LWR & 06629 & 44250.44821 & 1.73 & 1.72\\
LWR & 06867 & 44276.30139 & 1.69 & 1.72\\
LWR & 06868 & 44276.35351 & 1.70 & 1.74\\
LWR & 10147 & 44678.99903 & 1.77 & 1.74\\
LWR & 11631 & 44875.49513 & 1.89 & 1.73\\ 
LWP & 17753 & 47996.70867 & 1.25 & 1.39\\
LWP & 17803 & 48004.97747 & 1.26 & 1.40\\
LWP & 19825 & 48311.47734 & 1.35 & 1.45\\ 
LWP & 21760 & 48577.28020 & 1.28 & 1.40\\ 
LWP & 21979 & 48599.49844 & 1.19 & 1.39\\ 
LWP & 21980 & 48599.78824 & 1.36 & 1.19\\ 
LWP & 22578 & 48693.92689 & 1.36 & 1.42\\ 
LWP & 23595 & 48831.55923 & 1.27 & 1.35\\ 
LWP & 24599 & 48982.92294 & 1.27 & 1.36\\
LWP & 26436 & 49252.33258 & 0.932& 1.07\\
LWP & 28072 & 49479.38554 & 1.33 & 1.44\\
LWP & 29749 & 49710.25362 & 1.27 & 1.38\\
\hline
\multicolumn{3}{l}{mean uncertainty} & 0.3 & 0.1\\
\hline\\ \\ \\ \\
\end{tabular}
\end{center}
\end{table}

\begin{table}
\caption{List of the IUE observations of HD 37974. The units of broad-band
fluxes
$F_{1500}$, $F_{2175}$, and $F_{2500}$ are $10^{-13}\,\ergsa$, the units of $F_\text{em}$ (\ion{O}{i} 1302\,\AA)
are $10^{-12}\,\ergs$, the units of $F_\text{em}$ (total) are $10^{-11}\,\ergs$,
and the units of $F_\text{ab}$ (total) are $10^{-11}\,\ergs$.
We have not used LWP spectra 17752 and 26022 for our analysis, because their
flux around 2000\,\AA\ does not correspond to the SWP spectra taken roughly at
the same time.}
\begin{center}
\begin{tabular}{ccccccc}
\hline
Camera & Image & Julian date & $F_{1500}$ & $F_\text{em}$ & $F_\text{em}$ &
$F_\text{ab}$\\
       &&   2,400,000.5+&&\ion{O}{i} 1302\,\AA & \multicolumn{2}{c}{total}\\
\hline
SWP & 01402 & 43620.84477 & 7.19 & 3.5 & 3.1 & 2.8\\
SWP & 16617 & 45052.96092 & 6.75 & 2.5 & 2.4 & 2.4\\
SWP & 21142 & 45600.65824 & 6.44 & 1.8 & 2.2 & 2.3\\
SWP & 36705 & 47726.43124 & 6.32 & 1.5 & 2.0 & 2.2\\
SWP & 38466 & 47979.76919 & 6.58 & 2.7 & 2.3 & 2.3\\
SWP & 44640 & 48752.84347 & 6.79 & 3.1 & 2.6 & 2.4\\
SWP & 46715 & 48999.78725 & 6.65 & 2.9 & 2.6 & 2.1\\
SWP & 48224 & 49192.51418 & 6.85 & \\
SWP & 53170 & 49710.35884 & 6.62 & 3.3 & 2.7 & 2.0\\
\hline
\multicolumn{3}{l}{mean uncertainty} & 0.4 & 0.6 & 0.2 & 0.2\\
\hline
\end{tabular}\qquad
\begin{tabular}{ccccc}
\hline
Camera & Image & Julian date & $F_{2175}$ & $F_{2500}$\\
       &&   2,400,000.5+\\
\hline
LWR & 01364 & 43620.77220 & 5.28 & 4.89\\
LWR & 01365 & 43620.81664 & 5.67 & 6.46\\
LWR & 01366 & 43620.88505 & 5.86 & 6.08\\
LWP & 02005 & 45600.74598 & 3.94 & 5.00\\
LWR & 12854 & 45052.94009 & 5.20 & 5.83\\
LWP & 15959 & 47726.40155 & 5.10 & 5.62\\
LWP & 23083 & 48752.87841 & 5.27 & 5.69\\
LWP & 24721 & 48999.80857 & 4.95 & 5.55\\
\hline
\multicolumn{3}{l}{mean uncertainty} & 0.8 & 0.5\\
\hline\\
\end{tabular}
\end{center}
\end{table}

\begin{table}
\caption{List of the IUE observations of HD 38489. The units of broad-band
fluxes
$F_{1500}$, $F_{2175}$, and $F_{2500}$ are $10^{-13}\,\ergsa$, the units of
$F_\text{em}$ (\ion{He}{ii} 1640\,\AA, \ion{O}{iii} 1665\,\AA, and \ion{N}{iii}
1752\,\AA)
are $10^{-12}\,\ergs$, the units of $F_\text{em}$ (total) are $10^{-11}\,\ergs$,
and the units of $F_\text{ab}$ (total) are $10^{-12}\,\ergs$.}
\label{hd38489tab}
\begin{center}
\begin{tabular}{ccccccccc}
\hline
Camera & Image & Julian date & $F_{1500}$ & $F_\text{em}$ & $F_\text{em}$ &
$F_\text{em}$ &$F_\text{em}$ &$F_\text{ab}$\\
       &&   2,400,000.5+ &&\ion{He}{ii} 1640\,\AA&\ion{O}{iii} 1665\,\AA&
       \ion{N}{iii} 1752\,\AA & \multicolumn{2}{c}{total}\\
\hline
SWP & 14447 & 44796.61089 & 3.00 & 1.1 & 2.8 & 3.1 & 2.0 & 4.2\\
SWP & 15117 & 44875.62457 & 3.02 & 1.0 & 2.0 & 3.1 & 1.5 & 4.8\\
SWP & 25790 & 46182.62654 & 2.93 & 0.97& 1.8 & 2.4 & 1.7 & 4.5\\
SWP & 38464 & 47979.62427 & 2.91 & 1.7 & 2.5 & 2.5 & 1.9 & 4.7\\
SWP & 44163 & 48694.08735 & 2.86 & 1.5 & 2.6 & 2.9 & 2.0 & 3.0\\
SWP & 46714 & 48999.73408 & 3.06 & 1.7 & 2.3 & 3.0 & 2.2 & 3.3\\
SWP & 48223 & 49192.46020 & 3.06 & 1.3 & 2.4 & 3.1 & 2.0 & 4.4\\
\hline
\multicolumn{3}{l}{mean uncertainty} & 0.2 & 0.4 & 0.3 & 0.4 & 0.2 & 0.4\\
\hline
\end{tabular}\qquad
\begin{tabular}{ccccc}
\hline
Camera & Image & Julian date & $F_{2175}$ & $F_{2500}$\\
       &&   2,400,000.5+ \\
\hline
LWR & 11047 & 44796.58758 & 1.47 & 1.67\\
LWR & 11634 & 44875.64559 & 1.52 & 1.73\\
LWP & 17644 & 47979.64906 & 1.47 & 1.69\\
LWP & 22580 & 48694.05967 & 1.47 & 1.68\\
LWP & 22581 & 48694.11313 & 1.50 & 1.74\\
LWP & 24720 & 48999.75894 & 1.51 & 1.70\\
LWP & 25987 & 49192.47408 & 1.52 & 1.70\\
\hline
\multicolumn{3}{l}{mean uncertainty} & 0.4 & 0.2 \\
\hline
\end{tabular}
\end{center}
\end{table}

\begin{table}
\caption{List of the IUE observations of HD 45677. The units of broad-band
fluxes
$F_{1500}$, $F_{2175}$, and $F_{2500}$ are $10^{-12}\,\ergsa$, the units of
$F_\text{em}$ (\ion{O}{i} 1641\AA)
are $10^{-12}\,\ergs$, the units of $F_\text{em}$ (\ion{N}{i} 1745\,\AA) are
$10^{-11}\,\ergs$, and the units of $F_\text{em}$ (total) and 
$F_\text{ab}$ (total) are $10^{-10}\,\ergs$.}
\label{hd45677tab}
\begin{center} 
\begin{tabular}{cccccccc}
\hline
Camera & Image & Julian date & $F_{1500}$ & 
$F_\text{em}$ &$F_\text{em}$ &$F_\text{em}$ & $F_\text{ab}$\\
       &&   2,400,000.5+&&\ion{O}{i}&\ion{N}{i} & \multicolumn{2}{c}{total}\\
\hline
SWP & 01388 & 43617.91900 & 7.98 &\\
SWP & 02707 & 43771.73863 & 8.62 &\\
SWP & 02772 & 43777.64468 & 8.86 &\\
SWP & 04340 & 43926.86856 &      &     & 1.5& 2.4 & 1.8\\
SWP & 04761 & 43959.28757 &      &     & 1.4& 2.6 & 1.6\\
SWP & 06569 & 44135.58933 & 7.78 &\\  
SWP & 08006 & 44290.99348 & 8.14 &\\ 
SWP & 10648 & 44564.23906 & 8.18 &\\
SWP & 10652 & 44564.46919 &      & 20  & 1.2& 2.6 & 1.1\\
SWP & 11230 & 44639.19581 & 8.14 &\\
SWP & 15991 & 44978.02437 &      & 14  & 1.3& 3.2 & 1.4\\
SWP & 15992 & 44978.12183 & 8.86 &\\
SWP & 40272 & 48230.26057 &      & 24  & 1.6& 3.6 & 1.9\\
SWP & 44033 & 48673.17445 & 14.4 &\\
SWP & 44069 & 48679.02518 & 13.6 &\\
SWP & 46547 & 48978.12535 &      & 28  & 1.9& 3.9 & 2.2\\
SWP & 46548 & 48978.23332 &      & 33  & 1.5& 4.1 & 2.1\\
SWP & 46549 & 48978.34512 & 12.7 &\\
SWP & 48869 & 49268.38414 &      & 34  & 1.7& 4.8 & 1.8\\
SWP & 50020 & 49397.96430 &      & 40  & 1.7& 5.6 & 2.0\\
SWP & 50021 & 49398.00809 & 12.5 &\\
SWP & 50042 & 49400.11357 & 11.6 &\\
SWP & 52074 & 49605.35304 &      & 33  & 2.0& 4.7 & 2.2\\
SWP & 52075 & 49605.40010 & 12.6 &\\
SWP & 53046 & 49698.45771 &      & 47  & 1.9& 4.8 & 3.1\\
SWP & 55972 & 49980.35381 &      & 35  & 1.5& 3.7 & 1.9\\
SWP & 55973 & 49980.40276 & 13.5 &\\
\hline
\multicolumn{3}{l}{mean uncertainty} & 0.5 & 1.8 & 0.2 & 0.2 & 0.1\\
\hline
\end{tabular}\qquad
\begin{tabular}{ccccc}
\hline
Camera & Image & Julian date & $F_{2175}$ & $F_{2500}$\\
       &&   2,400,000.5+\\
\hline
LWR & 01342 & 43617.86552 & 3.88 & 4.43\\
LWR & 01343 & 43617.95139 & 3.85 & 4.48\\
LWR & 02416 & 43771.69562 & 4.08 & 4.77\\
LWR & 02467 & 43777.63777 & 4.19 & 5.09\\
LWR & 05630 & 44135.58619 & 3.87 & 4.67\\
LWR & 06966 & 44290.98922 & 3.69 & 4.46\\
LWR & 09357 & 44564.23236 & 4.00 & 4.51\\
LWR & 09850 & 44639.18781 & 3.73 & 4.65\\
LWR & 12309 & 44978.12974 & 4.28 & 4.86\\
LWP & 22428 & 48673.16886 & 6.37 & 5.92\\
LWP & 22467 & 48679.00908 & 6.29 & 7.14\\
LWP & 24554 & 48978.34880 & 6.11 & 6.83\\
LWP & 27420 & 49397.98586 & 5.81 & 6.57\\
LWP & 29140 & 49605.39065 & 5.93 & 6.22\\
LWP & 31508 & 49980.39867 & 5.82 & 6.79\\
\hline
\multicolumn{3}{l}{mean uncertainty} & 0.7 & 0.4 \\
\hline\\\\\\\\\\\\\\\\\\\\\\\\
\end{tabular}
\end{center}
\end{table}

\begin{table}
\caption{List of the IUE observations of HD 50138. The units of broad-band
fluxes
$F_{1500}$, $F_{2175}$, and $F_{2500}$ are $10^{-11}\,\ergsa$, the units of
$F_\text{em}$ (1530\,\AA)
are $10^{-11}\,\ergs$, and the units of $F_\text{em}$ and
$F_\text{ab}$ (total) are $10^{-10}\,\ergs$.}
\label{hd50138tab}
\begin{center}
\begin{tabular}{ccccccc}
\hline
Camera & Image & Julian date & $F_{1500}$ &
$F_\text{em}$ &$F_\text{em}$ &$F_\text{ab}$\\
       &&   2,400,000.5+&& 1530\,\AA & \multicolumn{2}{c}{total}\\
\hline
SWP & 03663 & 43863.62772  & 2.19 & \\
SWP & 04294 & 43922.93508  & 1.92 & \\
SWP & 04295 & 43923.02686  & 1.81 & 4.9 & 13.2  & 3.2\\
SWP & 04557 & 43941.29904  & 2.36 & 5.8 & 10.5 & 5.9\\
SWP & 04947 & 43978.87895  & 2.40 \\
SWP & 08007 & 44291.06204  & 2.08 & 3.2 & 10.1 & 4.7\\
SWP & 08008 & 44291.10022  & 2.20 \\
SWP & 10649 & 44564.30298  & 1.40 & 4.1 & 8.6  & 4.4\\
SWP & 10650 & 44564.33850  & 1.46 \\
SWP & 11231 & 44639.24050  & 1.83 \\
SWP & 15993 & 44978.17375  & 1.75 & 2.9 & 9.9  & 3.2\\
SWP & 15994 & 44978.22819  & 1.89 \\
SWP & 16017 & 44980.17490  & 1.80 & 3.3 & 8.5  & 4.9\\
SWP & 44190 & 48699.66053  & 1.88 & 4.7 & 10.7 & 6.5\\
SWP & 50022 & 49398.06495  & 2.81 \\
SWP & 50023 & 49398.12977  & 2.61 & 4.1 & 10.2 & 5.7\\
SWP & 50024 & 49398.16796  & 2.87 \\
SWP & 52169 & 49616.47446  & 2.62 & 9.0 & 12.0 & 6.3\\
SWP & 53047 & 49698.53272  & 2.25 & 6.3 & 12.7 & 4.8\\
SWP & 54035 & 49780.85206  & 2.10 & 4.4 & 13.8 & 5.5\\
\hline
\multicolumn{3}{l}{mean uncertainty} & 0.09 & 0.5 & 0.5 & 0.2\\
\hline\\\\\\\\\\\\\\\\\\\\\\\\\\\\\\\\\\\\\\\\\\\\
\end{tabular}\qquad
\begin{tabular}{ccccc}
\hline
Camera & Image & Julian date & $F_{2175}$ & $F_{2500}$\\
       &&   2,400,000.5+\\
\hline
LWR & 02499 & 43780.64124  & 1.50  & 1.26 \\
LWR & 03226 & 43863.52685  & 1.35  & 1.11 \\
LWR & 03247 & 43865.44774  & 1.44  & 1.31 \\
LWR & 03794 & 43922.92648  & 1.17  & 1.04 \\
LWR & 03796 & 43923.00149  & 1.17  & 1.05 \\
LWR & 03967 & 43941.27313  & 1.47  & 1.24 \\
LWR & 04276 & 43978.86907  & 1.37  & 1.14 \\
LWR & 06958 & 44289.23593  & 1.09  & 0.847\\
LWR & 06967 & 44291.02461  & 1.09  & 0.929\\
LWR & 06968 & 44291.09541  & 1.21  & 0.985\\
LWR & 09358 & 44564.27408  & 0.585 & 0.574\\
LWR & 09359 & 44564.33405  & 0.789 & 0.680\\
LWR & 09851 & 44639.23849  & 0.908 & 0.724\\
LWR & 12310 & 44978.19663  & 0.814 & 0.738\\
LWR & 12311 & 44978.23478  & 0.962 & 0.821\\
LWR & 12322 & 44980.15007  & 0.894 & 0.740\\
LWP & 21563 & 48559.04685  & 0.980 & 0.801\\
LWP & 22631 & 48699.68528  & 1.02  & 0.824\\
LWP & 27421 & 49398.06091  & 1.50  & 1.37 \\
LWP & 27422 & 49398.09727  & 1.46  & 1.26 \\
LWP & 27423 & 49398.14733  & 1.53  & 1.33 \\
LWP & 27430 & 49400.06514  & 1.33  & 1.19 \\
LWP & 29213 & 49616.44913  & 1.43  & 1.11 \\
LWP & 29691 & 49698.49794  & 1.24  & 1.12 \\
LWP & 30154 & 49780.82694  & 1.19  & 1.01 \\
LWP & 30212 & 49787.84814  & 0.995 & 0.942\\
LWP & 30215 & 49788.91653  & 1.09  & 0.989\\
LWP & 30230 & 49790.05439  & 1.19  & 0.986\\
LWP & 30236 & 49791.10010  & 1.22  & 1.01 \\
LWP & 30243 & 49792.04370  & 1.28  & 1.07 \\
LWP & 30248 & 49792.82657  & 1.13  & 0.942\\
LWP & 30251 & 49793.82873  & 1.02  & 0.880\\
LWP & 30256 & 49794.95347  & 1.15  & 1.04 \\
LWP & 30261 & 49795.81162  & 1.17  & 1.01 \\
LWP & 30268 & 49796.82841  & 1.32  & 1.06 \\
LWP & 30280 & 49797.91027  & 1.20  & 1.08 \\
LWP & 30290 & 49799.10282  & 1.46  & 1.26 \\
LWP & 30297 & 49800.10454  & 1.21  & 0.970\\
LWP & 30299 & 49800.93271  & 1.16  & 0.933\\
LWP & 30305 & 49801.80813  & 1.30  & 1.06 \\
LWP & 30311 & 49802.83435  & 1.35  & 1.10 \\
LWP & 30316 & 49803.82211  & 1.24  & 1.06 \\
\hline
\multicolumn{3}{l}{mean uncertainty} & 0.2 & 0.07 \\
\hline 
\end{tabular}
\end{center}
\end{table}


\begin{table}
\caption{List of the IUE observations of HD 87643. The units of broad-band
fluxes
$F_{1500}$, $F_{2175}$, and $F_{2500}$ are $10^{-13}\,\ergsa$,
the units of $F_\text{em}$ ($1730$\,\AA) are $10^{-12}\,\ergs$, the units of
$F_\text{em}$ (total) are $10^{-11}\,\ergs$,  and
$F_\text{ab}$ (total) are $10^{-12}\,\ergs$.}
\label{slephd87643}
\begin{center}
\begin{tabular}{ccccccc}
\hline
Camera & Image & Julian date & $F_{1500}$ &
$F_\text{em}$ &$F_\text{em}$ &$F_\text{ab}$\\
       &&   2,400,000.5+ && $1730$\,\AA & \multicolumn{2}{c}{total}\\
\hline
SWP & 05838 & 44071.98731 & 3.56 & 2.9 & 5.0 & 2.9\\
SWP & 05889 & 44076.10701 & 3.53 & 3.0 & 4.2 & 7.7\\
SWP & 09166 & 44391.17960 & 3.83 &     &     & \\
SWP & 13760 & 44714.20766 & 3.48 & 2.6 & 4.3 & 3.5\\
SWP & 25996 & 46208.58548 & 3.07 & 2.4 & 3.5 & 3.1\\
SWP & 26040 & 46216.42460 & 3.54 & 2.5 & 4.0 & 4.8\\
SWP & 28309 & 46564.78045 & 2.43 & 2.6 & 4.2 & 4.8\\
SWP & 36803 & 47744.58941 & 4.01 & 2.4 & 4.3 & 6.1\\
SWP & 46913 & 49029.08163 & 3.09 & 2.2 & 3.4 & 5.2\\
SWP & 50710 & 49479.33012 & 2.21 & 1.9 & 2.4 & 5.0\\
SWP & 53842 & 49754.11026 & 2.17 & 2.1 & 2.5 & 4.6\\
\hline
\multicolumn{3}{l}{mean uncertainty} & 0.2 & 0.3 & 0.4 & 0.5\\
\hline\\\\\\
\end{tabular}\qquad
\begin{tabular}{ccccc}
\hline
Camera & Image & Julian date & $F_{2175}$ & $F_{2500}$\\
       &&   2,400,000.5+\\
\hline
LWR & 05084 & 44071.96524 & 0.990 & 1.83\\
LWR & 05085 & 44072.00495 & 1.00  & 1.87\\
LWR & 05144 & 44076.09434 & 1.12  & 1.91\\
LWP & 06038 & 46208.60380 & 0.933 & 2.20\\
LWP & 06098 & 46216.40606 & 0.911 & 2.30\\
LWP & 08199 & 46564.80108 & 0.697 & 1.39\\
LWR & 10392 & 44714.19108 & 1.13  & 2.68\\
LWR & 10393 & 44714.31543 & 1.27  & 2.94\\
LWP & 16078 & 47744.57379 & 0.938 & 2.54\\
LWP & 20595 & 48421.32603 & 0.298 & 1.18\\
LWP & 24888 & 49028.92607 & 0.601 & 2.30\\
LWP & 24890 & 49029.17346 & 0.705 & 1.87\\
LWP & 26086 & 49204.77187 & 0.596 & 1.68\\
LWP & 26143 & 49213.18713 & 0.739 & 1.79\\
\hline
\multicolumn{3}{l}{mean uncertainty} & 0.9 & 0.3 \\
\hline
\end{tabular}
\end{center}
\end{table}

\begin{table}
\caption{List of the IUE observations of HD 94878. The units of broad-band
fluxes
$F_{1500}$, $F_{2175}$, and $F_{2500}$ are $10^{-13}\,\ergsa$, the units of
$F_\text{ab}$ (\ion{C}{ii} 1335\,\AA\ and \ion{C}{iv} 1548\,\AA)
are $10^{-12}\,\ergs$, the units of $F_\text{em}$ and
$F_\text{ab}$ (total) are $10^{-11}\,\ergs$.}
\label{slephd94878}
\begin{center}
\begin{tabular}{cccccccc}
\hline
Camera & Image & Julian date & $F_{1500}$ &
$F_\text{ab}$ &$F_\text{ab}$ &$F_\text{em}$ &$F_\text{ab}$\\
       &&   2,400,000.5+&&\ion{C}{ii}&\ion{C}{iv} & \multicolumn{2}{c}{total}\\
\hline
SWP & 02685 & 43769.84366  & 19.4 &     & 9.2\\
SWP & 02686 & 43769.89388  & 24.2 & 8.8 & 9.6 & 7.1 & 7.2\\
SWP & 05679 & 44054.14513  & 26.5 & 3.5 & 13  & 4.9 & 8.8\\
SWP & 06737 & 44149.71582  & 21.2 & 3.3 & 7.8 & 3.4 & 6.5\\
SWP & 08936 & 44365.16023  & 19.6\\
SWP & 09309 & 44408.86859  & 26.1 & 7.2 & 11  & 4.3 & 7.3\\
SWP & 18555 & 45287.70249  & 23.1 & 7.4 & 12  & 3.7 & 8.1\\
\hline
\multicolumn{3}{l}{mean uncertainty} & 1.1 & 1.6 & 2 & 0.3 & 0.5\\
\hline
\end{tabular}\qquad
\begin{tabular}{ccccc}
\hline
Camera & Image & Julian date & $F_{2175}$ & $F_{2500}$\\
       &&   2,400,000.5+\\
\hline
LWR & 02396 & 43769.77632  & 5.41 & 9.75\\
LWR & 02397 & 43769.86043  & 4.92 & 12.1\\
LWR & 04921 & 44054.11469  & 5.09 & 12.5\\
LWR & 07683 & 44365.30913  & 3.69 & 8.75\\
LWR & 08071 & 44408.88296  & 5.41 & 10.9\\
LWR & 14625 & 45287.68397  & 4.27 & 9.09\\
\hline
\multicolumn{3}{l}{mean uncertainty} & 1.2 & 1.0 \\
\hline\\
\end{tabular}
\end{center}
\end{table}

\begin{table}
\caption{List of the IUE observations of HD 100546. The units of broad-band
fluxes
$F_{1500}$, $F_{2175}$, and $F_{2500}$ are $10^{-11}\,\ergsa$, the units of
$F_\text{ab}$ (\ion{C}{ii} 1335\,\AA\ and \ion{Fe}{ii} 1657\,\AA)
are $10^{-11}\,\ergs$, the units of $F_\text{em}$ (1779\,\AA\ + 1784\,\AA) are
$10^{-11}\,\ergs$, and
 the units of $F_\text{em}$ and
$F_\text{ab}$ (total) are $10^{-10}\,\ergs$.}
\label{slephd100546}
\begin{center}
\begin{tabular}{ccccccccc}
\hline
Camera & Image & Julian date& $F_{1500}$ &
$F_\text{ab}$ &$F_\text{ab}$ &$F_\text{em}$ &$F_\text{em}$ & $F_\text{ab}$ \\
       &&   2,400,000.5+ & & \ion{C}{ii} & \ion{Fe}{ii}& 1779\,\AA+
       & \multicolumn{2}{c}{total}\\
\hline
SWP & 39712 & 48160.68056 & 2.15 & 5.2 & 1.9 & 0.76 & 1.9 & 4.5\\
SWP & 47481 & 49091.17277 & 2.07 & 5.1 & 0.75& 0.96 & 2.6 & 3.1\\
SWP & 54065 & 49783.82326 & 2.05 & 5.4 & 3.1 & 1.1  & 2.7 & 5.0 \\
SWP & 54658 & 49849.02155 & 2.26 &    \\
SWP & 54753 & 49862.71465 & 2.38 &    \\
SWP & 54754 & 49862.79287 & 2.18 & 5.1 & 1.2 & 1.2  & 3.4 & 3.3\\
\hline
\multicolumn{3}{l}{mean uncertainty} & 0.07 & 0.5 & 0.3 & 0.5 & 0.1 & 0.2\\
\hline\\
\end{tabular}
\begin{tabular}{ccccc}
\hline
Camera & Image & Julian date & $F_{2175}$ & $F_{2500}$\\
       &&   2,400,000.5+\\
\hline
LWP & 16052 & 47740.75450 & 1.06  & 0.913\\
LWP & 16075 & 47744.03140 & 1.06  & 0.904\\
LWP & 30679 & 49849.01533 & 1.14  & 1.00 \\
LWP & 30770 & 49862.70994 & 1.12  & 0.959\\
LWP & 30771 & 49862.76354 & 0.984 & 0.842\\
\hline
\multicolumn{3}{l}{mean uncertainty} & 0.2 & 0.06 \\
\hline\\\\
\end{tabular}
\end{center}
\end{table}

\begin{table}
\caption{List of the IUE observations of HD 169515. The units of broad-band
fluxes
$F_{1500}$, $F_{2175}$, and $F_{2500}$ are $10^{-14}\,\ergsa$, the units of
$F_\text{em}$ (\ion{C}{iv} $1550\,$\AA)
are $10^{-13}\,\ergs$, the units of $F_\text{ab}$ (\ion{Fe}{ii} 1658\,\AA) are
 $10^{-13}\,\ergs$, and the units of $F_\text{em}$ and
$F_\text{ab}$ (total) are $10^{-12}\,\ergs$.}
\label{slephd169515}
\begin{center}
\begin{tabular}{ccccccccc}
\hline
Camera & Image & Julian date& $F_{1500}$ &
$F_\text{em}$ &$F_\text{ab}$ &$F_\text{em}$ & $F_\text{ab}$ \\
       &&   2,400,000.5+&&\ion{C}{iv}&\ion{Fe}{ii} & \multicolumn{2}{c}{total}\\
\hline
SWP & 01543 & 43642.75860 & 8.14 & 3.0 & 2.3 & 10.2 & 5.1\\
SWP & 05830 & 44071.39234 & 8.71 & 4.8 & 0.9 & 4.9  & 2.6\\
SWP & 05831 & 44071.47761 & 10.6 & 4.6 & 1.0 & 6.3  & 2.6\\
SWP & 06614 & 44139.72653 & 13.0 & 7.7 & 3.1 & 4.4  & 2.5\\
SWP & 08938 & 44365.53453 & 7.90 & 6.9 & 1.4 & 3.4  & 1.0\\
SWP & 17386 & 45159.00706 & 12.7 & 8.9 & 2.8 & 4.5  & 1.7\\
SWP & 17408 & 45161.95042 & 10.5 & 5.9 & 2.2 & 2.3  & 2.0\\
SWP & 18099 & 45236.89806 & 12.4 & 6.2 & 2.9 & 3.4  & 2.4\\
SWP & 33431 & 47283.76895 & 13.0 & 5.7 & 2.7 & 3.0  & 2.1\\
\hline
\multicolumn{3}{l}{mean uncertainty} & 1.5 & 3 & 0.9 & 0.6 & 0.3\\
\hline\\\\
\end{tabular}\qquad
\begin{tabular}{ccccc}
\hline
Camera & Image & Julian date & $F_{2175}$ & $F_{2500}$ \\
       &&   2,400,000.5+\\
\hline
LWR & 01493 & 43642.72896  & 0.706 & 5.30\\
LWR & 05078 & 44071.38159  & 0.172 & 6.19\\
LWR & 05079 & 44071.41798  & 0.456 & 6.68\\
LWR & 05672 & 44139.69677  & 0.594 & 8.16\\
LWR & 05673 & 44139.74953  & 0.528 & 8.02\\
LWR & 07686 & 44365.56572  & 0.175 & 4.51\\
LWP & 13161 & 47283.80280  & 0.753 & 8.02\\
LWR & 13636 & 45158.93463  & 0.572 & 7.52\\
LWR & 13658 & 45161.90118  & 0.426 & 6.30\\
LWR & 13659 & 45161.99969  & 0.634 & 6.56\\
LWR & 14249 & 45236.92795  & 0.643 & 7.63\\
\hline
\multicolumn{3}{l}{mean uncertainty} & 1.7 & 1.2 \\
\hline
\end{tabular}
\end{center}
\end{table}

\bsp    
\label{lastpage}
\end{document}